\documentclass[journal]{IEEEtran}
\pdfoutput=1
\usepackage{todonotes}

\usepackage{epstopdf}

\DeclareFontFamily{OT1}{pzc}{}
\DeclareFontShape{OT1}{pzc}{m}{it}{<-> s * [1.500] pzcmi7t}{}
\DeclareMathAlphabet{\mathpzc}{OT1}{pzc}{m}{it}

\usepackage{multicol}

\usepackage{amsmath}
\usepackage{array}

\ifCLASSOPTIONcompsoc
 \usepackage[caption=false,font=normalsize,labelfont=sf,textfont=sf]{subfig}
\else
 \usepackage[caption=false,font=footnotesize]{subfig}
\usepackage{lipsum}

\usepackage{cuted}

\usepackage{fixltx2e}

\usepackage{stfloats}
\begin{document}
%
\title{Generalized group-based epidemic model for spreading processes on networks: GgroupEM}
%
%
%

\author{Sifat~Afroj~Moon,~\IEEEmembership{Member,~IEEE,}
Faryad~Darabi~Sahneh,~\IEEEmembership{Member,~IEEE,}
Caterina~Scoglio,~\IEEEmembership{Member,~IEEE}
\thanks{Sifat Afroj Moon and Caterina Scoglio are with the Department
of Electrical and Computer Engineering, Kansas State University, Kansas,
KS, 66502 USA e-mail: \{sifatafroj,caterina\}@k-state.edu.}
\thanks{Faryad Sahneh is with the Department of Computer Science at University of Arizona, Tucson, AZ 85721. e-mail: faryad@cs.arizona.edu}
\thanks{Manuscript received ; revised .}}

%
%

\markboth{Journal of \LaTeX\ Class Files,~Vol.~14, No.~8, August~2015}%
{Shell \MakeLowercase{\textit{et al.}}: Bare Demo of IEEEtran.cls for IEEE Journals}
%



\maketitle

\begin{abstract}
We develop a generalized group-based epidemic model (GgroupEM) framework for any compartmental epidemic model (for example; susceptible-infected-susceptible, susceptible-infected-recovered, susceptible-exposed-infected-recovered). Here, a group consists of a collection of individual nodes. This model can be used to understand the important dynamic characteristics of a stochastic epidemic spreading over very large complex networks, being informative about the state of groups. Aggregating nodes by groups, the state space becomes smaller than the individual-based approach at the cost of aggregation error, which is strongly bounded by the isoperimetric inequality. We also develop a mean-field approximation of this framework to further reduce the state-space size. Finally, we extend the GgroupEM to multilayer networks. Since the group-based framework is computationally less expensive and faster than an individual-based framework, then this framework is useful when the simulation time is important.

\end{abstract}

\begin{IEEEkeywords}
compartmental model, epidemic model, continuous-time Markov process, mean-field approximation, network, spreading process, scaling, graph partitioning.
\end{IEEEkeywords}

%
\IEEEpeerreviewmaketitle


\section{Introduction}
%
%
%
%
\IEEEPARstart{E}{pidemic} spreading processes over complex networks is an important topic for different research fields, such as epidemiology, social science, computer science, etc. \cite{barabasi2016network, vespignani2012modelling, pastor2015epidemic, moon2019estimation, ferdousi2019understanding, barrat2008dynamical}. The theoretical model of stochastic epidemic spreading processes over a network can reveal important dynamic characteristics of the epidemic. The spread of computer virus, information, opinions, rumors, knowledge, products, or any spreading process in a network of interactive agents can be modeled as the epidemic process. All the above spreading processes follow some common patterns.
\\
Compartmental models are widely used in the study of epidemics. In a compartmental model, individuals/agents can be in different compartments. Generally, compartments represent the fraction of a homogeneous population in a given state. The set of compartments can be different for different models. The widely used compartments in the literature are susceptible, infected, recovered, immune, and latent \cite{barabasi2016network}. The compartments can be different for different research areas or scenarios. An individual can move from one compartment to another compartment when this event is assumed to be an independent Poisson process with a constant rate, this assumption leads to a continuous-time Markov process.  \\
Some complex networks have a large set of nodes/agents. Modeling epidemic spreading over those very large networks is computationally expensive and time-consuming. To address this issue, the group-based epidemic model can be useful. Sahneh et. al. proposed a generalized epidemic model framework (GEMF) for the individual-based approach \cite{sahneh2013generalized}. This individual-based approach can estimate the state of any individual node or agent at any time $t$. However, for some scenarios, it is redundant to know the state of every node, as in the case of a network of farm animals that contains millions of farm animals. In this scenario, the individual-based approach is not only computationally expensive but also not required. To face this complexity, sometimes researchers scale their network by reducing the number of nodes. To do that, they consider several individuals or a group as a single node \cite{mokross2014decay, white2017using, moon2019spatio}. This type of scaling can alter the actual network, and estimation of the dynamics of the system can be misleading. The group-based approach is a solution to this problem. A group can consist of farm animals of the same category, and the group state can tell us the summary of its node states. A group-based framework is not only useful in the network of livestock but also be useful to find out the dynamics of any type of network; such as, communication network, trade network, human network, biological network, power-grid network. In this new era, because of the improvement of the digital technology, different types of communication among humans are popular, as a result, a large number of people are connected to form very large networks. These networks can influence public opinion, which is very impactful in the field of politics, economy, business, and many other fields. A group-based framework can be useful to handle these very large networks to understand the different dynamics of its groups. It can also help us to find out the impact of the groups in the dynamics of the system.\\
The $N$-intertwined mean-field method (NIMFA) \cite{chakrabarti2008epidemic, van2009virus} and the heterogeneous mean-field method (HMF) \cite{pastor2001epidemic, boguna2002epidemic} are two well-established approximation methods for the analysis of dynamical processes on complex heterogeneous networks. They are two particular cases of the group-based approach unified mean-field framework (UMFF) which was first proposed by Devriendt et. al. for susceptible-infected-susceptible epidemic model \cite{devriendt2017unified}. The group-based approach has fewer degrees of freedom than NIMFA. Although HMF also has this property, however UMFF has more flexibility to choose groups than HMF. The heterogeneous mean-field method (HMF) is a degree based approach, and nodes of the same degree are assumed statistically equivalent, which is not the only case for UMFF.\\
In this paper, we generalize the group-based approach UMFF for any compartmental epidemic model. Generalization can increase its flexibility, compatibility, and applicability. Different dynamics can be modeled with a different compartmental model. Even only in the epidemiology, some diseases dynamics are more suitable by susceptible-infected-susceptible (SIS)  model, some are more suitable by susceptible-infected-recovered (SIR) compartmental model, and so on. A general framework can give more flexibility to the researches to model an epidemic spreading in a network. To derive a group-based general model, at first we develop a continuous-time Markov model for multi-compartmental node dynamics. Secondly, we propose the mean-field approximation of the continuous-time group-based Markov model. Finally, we provide the multilayer extension of the multi-compartmental group-based framework. The GgroupEM framework has a lower computation complexity and faster simulation time in comparison with the individual-based GEMF because of the reduced state space size.\\
This paper is organized as follows, some backgrounds of our work are reviewed in section \ref{Background}. In section \ref{exactModel}, we propose a continuous-time Markov process for a general group-based framework. Then, we provide the mean-field approximation for this framework in section \ref{UMFF}. We also provide some simulation results in this section. In section \ref{Multilayerextension}, we provide the multilayer extension of the GgroupEM framework. In the end, we provide some concluding remarks of our work in section \ref{Conclusion}.

\section{Backgrounds}
Here, we present some well known compartmental epidemic models based on group-based approach and discuss about the epidemic modelling on a network.
\label{Background}
\subsection{Compartmental epidemic models}
Compartmental models can describe the epidemic spreading on a network $\mathcal{G}(N, E)$ \cite{keeling2011modeling}. Here, $N$ is the number of nodes and $E$ is the set of edges in the network $\mathcal{G}$. In this paper, we present a group-based framework for any compartmental epidemic model. There are two types of transitions between compartments; 1) nodal transitions or independent transitions and 2) edge transitions or dependent transitions \cite{sahneh2013generalized}. The nodal transition of a node only depends on the current state of the node. The edge transition of a node depends on the current state of a node and the state of neighboring nodes. Each edge transition has an influencer compartment. This compartment can be defined as the compartment of the neighboring nodes or state of the neighboring nodes which affects that edge transition. For example, in susceptible-infected-susceptible epidemic, the susceptible to infected edge transition of a susceptible node depends on its infected neighboring nodes. Therefore, infected compartment is the influencer compartment for this edge transition. \\
Even though our focus is in the group-based approach, however, those node-level transitions are important even in the group-based approach. The change of a group state happens because of those node-level transitions. We can call those transitions as events. When an event (nodal or edge transition) happens on a node of a group, the group state changes. \\
Some common epidemic compartmental models are described here:
\subsubsection{Susceptible-infected-susceptible (SIS)}
This model has two compartments, $m \in \{1, 2\}$: susceptible $(m=1)$ and infected $(m=2)$. A node in the network can be susceptible or can be infected. There are two transitions in this model: one is edge transition (susceptible to infected) and another is nodal transition (infected to susceptible). Susceptible to the infected transition of a node depends on the infected neighbors of that node. The infected compartment is the influencer compartment for the transition susceptible to infected. In GgroupEM, each group will have two types of nodes; susceptible and infected. The group state will tell how many nodes are in each compartment. Let a group be in a state where it has $S$ susceptible nodes and $I$ infected nodes. If one infected node changes its compartment to the susceptible compartment, then the group state will change to $S+1$ susceptible nodes and $I-1$ infected nodes.  
\subsubsection{Susceptible-infected-recovered (SIR)}
This model has three compartments, $m \in \{1, 2, 3\}$: susceptible $(m=1)$, infected $(m=2)$, and recovered $(m=3)$. Each group can have these three types of nodes. There are also two transitions in this model: one is edge transition (susceptible to infected) and another is nodal transition (infected to susceptible). In GgroupEM, each group will have three types of nodes: susceptible, infected, and recovered nodes.  
\subsubsection{Susceptible-exposed-infected-recovered (SEIR)}
This is a variation of the SIR model. This model has four compartments, $m \in \{1, 2, 3, 4\}$: susceptible $(m=1)$, exposed $(m=2)$, infected $(m=3)$, and recovered $(m=4)$. There are three transitions in this model: one is an edge transition (susceptible to exposed) and the other two are nodal transitions (exposed to infected and infected to recovered). For this model, the infected compartment is the influencer compartment for the susceptible to exposed edge transition.  

These are some basic widely used epidemic compartmental model. A compartmental model can have any number of compartments and any number of transitions. Compartment number and type can be different in the scenario of rumor spreading or computer virus spreading.

\subsection{Epidemic modelling on networks}
A continuous-time Markov chain can model an epidemic process on a network, where each transition between compartments is an independent Poisson process with constant transition rate \cite{van1981stochastic}. The assumption of the independent Poisson process makes the system memoryless. In the individual-based approach, nodes are at the individual level. Each node has a fixed number of possible state. The state of a node in a network at time $t$ defined as $n_i(t)\in{1, 2, 3,...........M}$. In the epidemic process, if a node can move from compartment $2$ to compartment $1$ in a $\Delta t$ time by a nodal transition with rate $\delta$, then the waiting time for this transition is exponentially distributed with rate $\delta$. So,
\begin{align}
\begin{split}
     &Pr[n_i(t+\Delta t)=1|n_i(t)= 2 ]=  \delta\Delta t + o(\Delta t)
\end{split}
\end{align}
Here, $o(\Delta t)$ is a function of higher-order terms of $\Delta t$.\\
For an edge transition of node $i$ from compartment $1$ to $2$ with a transition rate $\beta$, when the node has one infected neighbor, the infection process for the node $i$ is,
\begin{equation}
    Pr[n_i(t+\Delta t)=2|n_i(t)= 1 ]=  \beta\Delta t + o(\Delta t)
\end{equation}
In this work, we have assumed that each infective link can transmit the disease with constant rate $\beta$. In a network of $N$ nodes, the individual-based or node-based Markov chain has $M^N$ possible states for a $M$ compartmental epidemic model \cite{sahneh2013generalized}. \\
In the following section, we present the group-based continuous-time Markov chain epidemic modeling on a network. All the symbols and their definitions to develop this model are given in the Table \ref{table_list}.
\begin{table}[htb]
\renewcommand{\arraystretch}{1.5}
\caption{Notation of parameters}
\label{table_list}
\centering
\begin{tabular}{c c}
\hline
\hline
Symbol & Definition\\
\hline
\hline
$N$ & number of nodes\\
\hline
$E$ & set of edges\\
\hline
$m$ & index variable for compartments\\
\hline
$M$ & number of compartments in the epidemic model\\
\hline
$t$ & time\\
\hline
$n_i$ & state of the $i$ node \\
\hline
$C$ & no. of groups\\
\hline
$L_{ij}$& no. of links from group $i$ to group $j$\\
\hline
$\mathcal{N}_i$ & no. of nodes in group $i$\\
\hline
$A$& adjacency matrix (dimension is $N\times N$)\\
\hline
$\mathcal{A}_g$& group-based adjacency matrix (dimension is $C\times C$)\\
\hline
$e_i^k$& state indicator vector of a group $i$\\
\hline
$V_i$& full state matrix of a group $i$\\
\hline
$x_{i,m}$ & no. of nodes in compartment $m$ in a group $i$\\
\hline
$g_i(t)$ & state of a group $i$ at time $t$\\
\hline
$G(t)$ & network state at time $t$\\
\hline
$\otimes$ & kronecker product\\
\hline
$\circ$ & element wise multiplication or Hadamard product \\
\hline
$\theta_i$ & transition matrix for a group $i$\\
\hline
$\Delta_{i,\delta_q}$ & \textit{transition indication matrix} for group $i$ for a nodal\\& transition $\delta_q$\\
\hline
$\Delta_{i,\beta_q}$ & \textit{transition indication matrix} for group $i$ for an edge\\& transition $\beta_q$\\

\hline
$q_n$& no. of nodal transitions \\
\hline
$q_e$ & no. of edge transitions \\
\hline
$\Theta$ & network state transition matrix\\
\hline
$\rho_i$& fraction of the nodes in each compartment in group $i$\\
\hline
$Q$& node-level transition matrix\\
\hline
$\mathcal{L}$& no. of layers in the multilayer network\\
\hline
$A_{gl}$& group-based adjacency matrix for a layer $l$ in the\\& multilayer network\\
\hline
\end{tabular}
\end{table}


\section{A group-based epidemic model: GgroupEM}
\label{exactModel}
\begin{figure}[htbp]
    \centering
   \includegraphics[width=0.45\textwidth]{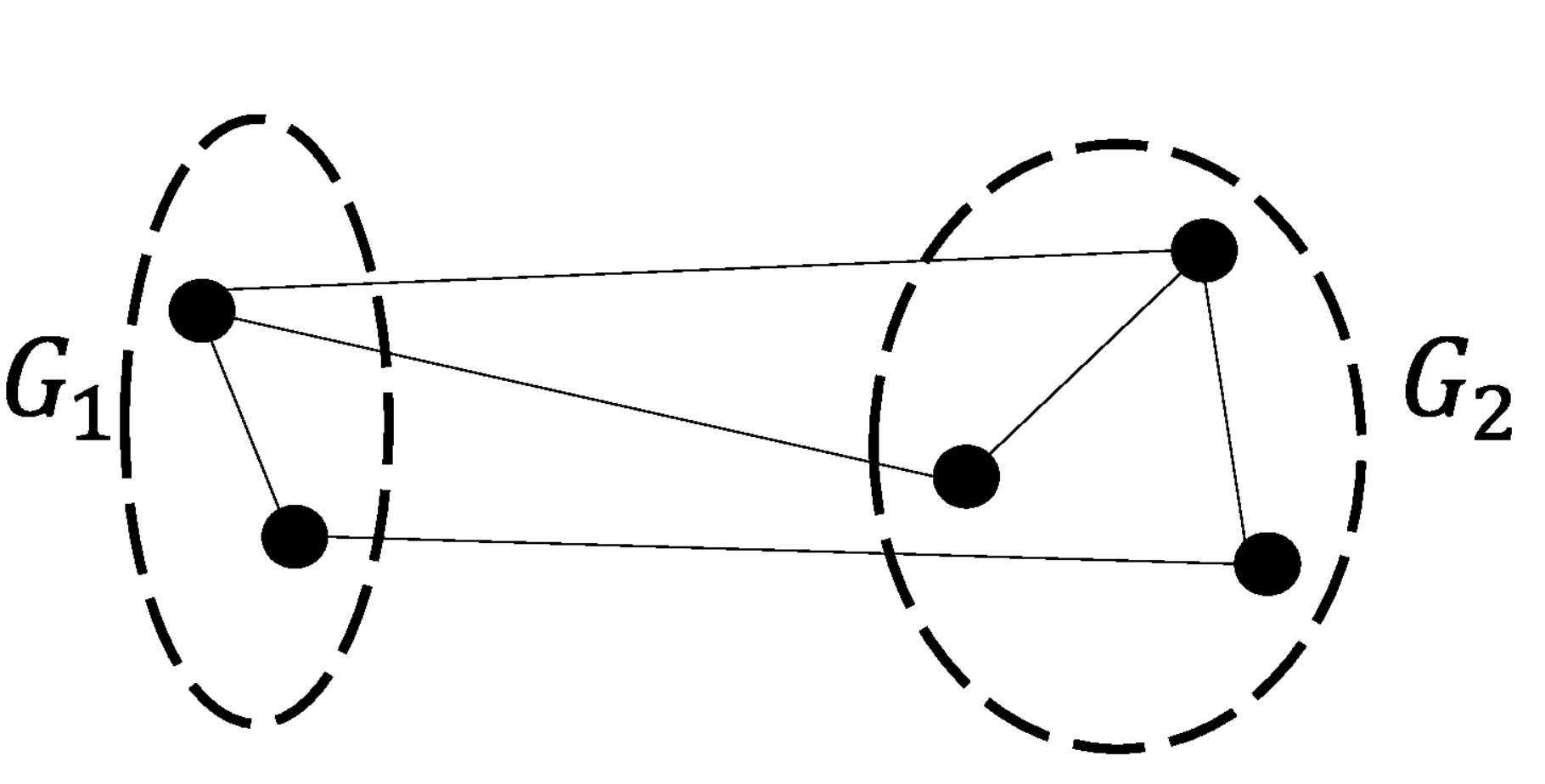}
  \caption{A group-based network. It has $N=5$ nodes, which are divided into two groups, $C=2$.}      
\label{fig1}
\end{figure}

In this model, the network consists of $N$ nodes, which are divided into $C$ disjoint nonempty groups. Where, 
\begin{equation}
    N= \mathcal{N}_1 + \mathcal{N}_2+.......+\mathcal{N}_C
\end{equation}
Here, $\mathcal{N}_i$ represents the number of nodes in a group $i$ and $i=1, 2, ..., C$. This group-based model does not contain information about each node state, it only contains the state of each group. The state of a group gives the number of nodes of that group in each compartment. The adjacency matrix $A$ of the network $\mathcal{G}$ is a $N\times N$ matrix, where each element is a binary number, \\
\begin{equation}
    A(i,j)=\begin{cases}1; &\text{if node $i$ to $j$ are connected by a link}\\
    0; & \text{otherwise}\end{cases}
\end{equation}
In undirected networks, $A(i,j)= A(j,i)$. The group-based adjacency matrix $\mathcal{A}_g$ is a $C \times C$ matrix. A element of the matrix $\mathcal{A}_g(i,j)$ represents the links from group $i$ to group $j$,
\begin{equation}
 \mathcal{A}_g(i,j)= \frac{\#\text{ of links from group $i$ to group $j$}}{\mathcal{N}_i\mathcal{N}_j}= \frac{L_{ij}}{\mathcal{N}_i\mathcal{N}_j} 
 \end{equation}
Here, $L_{ij}$ indicates number of the links from group $i$ to group $j$. 
\begin{equation}
    L_{ij}= u_iAu_j^T
    \label{linkeq}
\end{equation}
Where, $u_i$ is a $1\times N$ vector, where each element is $0$ or $1$. If $k$\textsuperscript{th} node is in group $i$, then $u_i(k)=1$ otherwise $u_i(k)=0$.\\
The group-based adjacency matrix $\mathcal{A}_g$ is a symmetric matrix for an undirected network. The diagonal elements of the $\mathcal{A}_g$ matrix is $\frac{L_{ii}}{(\mathcal{N}_i)^2}$, where $L_{ii}$ is the number of links inside the group, so $diag(\mathcal{A}_g)\geq 0$. For the bipartite networks, $diag(\mathcal{A}_g)= 0$. If $C==N$, then $\mathcal{A}_g= A$. An example of a group-based network are presented in Fig \ref{fig1}. The $\mathcal{A}_g$ matrix for this network is,
\[
\begin{bmatrix}
\frac{L_{11}}{\mathcal{N}_1\mathcal{N}_1}& \frac{L_{12}}{\mathcal{N}_1\mathcal{N}_2}\\
\frac{L_{21}}{\mathcal{N}_2\mathcal{N}_1}&\frac{L_{22}}{\mathcal{N}_2\mathcal{N}_2}\\
\end{bmatrix}=\begin{bmatrix}
\frac{1}{4}& \frac{3}{6}\\
\frac{3}{6}&\frac{2}{9}\\
\end{bmatrix}
\]
    
\subsection{Group state vectors}
For a $M$ compartmental epidemic model, the state vector of a group contains information about how many nodes are in each compartment. The number of possible states of a group with $\mathcal{N}_i$ nodes for the $M$ compartmental model is $\binom{\mathcal{N}_i +M-1}{M-1}$, which is determined from the \textit{stars and bars} combinatorics problem \cite{feller1957introduction}. This problem tells us the number of possible ways to put $\mathcal{N}_i$ indistinguishable nodes into $M$ distinguishable compartments. The state indicator vector of a group $i$ can be defined as,
\begin{equation}
    e_i^{k}:=[0, 0, 0.........\underbrace{1}_{\text{k\textsuperscript{th} entry}}....0]^T_{1\times\binom{\mathcal{N}_i +M-1}{M-1}}
\end{equation}
This state indicator vector indicates that group state is in $k$\textsuperscript{th} possible state.
The full state matrix $V_i$ of a group $i$ is a $\binom{\mathcal{N}_i +M-1}{M-1}\times M$ matrix. Each row of the matrix $V_i$ is, 
\begin{equation}
    [x_{i, 1}, x_{i, 2}, x_{i, 3}........x_{i, M}]
\end{equation}
Here, $x_{i, m}$ represents the number of nodes in the compartment $m$ in the group $i$, also $0\leq x_{i, 1}, x_{i, 2}, x_{i, 3}....x_{i, M}\leq \mathcal{N}_i$ and $\sum_{m=1}^{m=M}x_{i, m}= \mathcal{N}_i$. Let,
\begin{equation}
     X_i=[x_{i, 1}, x_{i, 2}, x_{i, 3}........x_{i, M}]^T = (V_i)^Te_i^k
\end{equation}
The $k$\textsuperscript{th} possible state can be obtained from the list of the all possible states. \\
In the following, we give two example of list of all the possible group states. \\
First example: $\mathcal{N}_i=2, M=3$, and number of possible state is $\binom{\mathcal{N}_i +M-1}{M-1}=6$. Possible states for this case are, 

\begin{equation}
 \overbrace{
\begin{bmatrix}
 | &|&o&o \\
 | &o&|&o \\
 | &o&o&| \\
 o &|&|&o \\
 o &|&o&| \\
 o &o&|&|
\end{bmatrix}}^{\text{Dividers and nodes}}
\quad \quad \quad \quad
\overbrace{
\begin{bmatrix} 
0&0&2\\
0&1&1\\
0&2&0\\
1&0&1\\
1&1&0\\
2&0&0
\end{bmatrix}}^{V_i}
\label{pattern1}
\end{equation}
Here, $|$ represents a divider and $o$ represents a node. Two dividers can divide the nodes into three compartments. The left matrix represent a chart of dividers and nodes. In each row, the nodes in the left side of the first divider are in the first compartment, the nodes in between the first divider and the second divider are in the second compartment and nodes in the right side of the second divider are in the third compartment. Each row represents a possible group state. First row represents that first compartment has zero nodes, second compartment also has zero nodes, and third compartment has two nodes. So, $x_{i,1}=0, x_{i,2}=0$, and $x_{i,3}=2$. The right matrix presents the full state matrix $V_i$, where first column is the number of nodes in the first compartment $x_{i,1}$, second column is the number of nodes in the second compartment ${x_{i,2}}$ and third column is the number of nodes in the third compartment $x_{i,3}$
\\Second example: $\mathcal{N}_i=3, M=3$, and number of possible state is $\binom{\mathcal{N}_i +M-1}{M-1}=10$,
\begin{equation}
\overbrace{
\begin{bmatrix}
 | &|&o&o&o\\
 | &o&|&o&o\\
| &o&o&|&o \\
 | &o&o&o&| \\
 o &|&|&o&o \\
 o &|&o&|&o \\
 o &|&o&o&| \\
 o &o&|&|&o \\ 
 o &o&|&o&| \\
 o &o&o&|&| 
\end{bmatrix}}^{\text{Dividers and nodes}}
\quad \quad \quad \quad
\overbrace{
\begin{bmatrix}
0&0&3\\
0&1&2\\
0&2&1\\
0&3&0\\
1&0&2\\
1&1&1\\
1&2&0\\
2&0&1\\
2&1&0\\
3&0&0\\
\end{bmatrix}}^{V_i}
\label{pattern2}
\end{equation}
 Here, we have two dividers to divide the three nodes into three compartments. We propose the pattern in expression (\ref{pattern1}) and (\ref{pattern2}) to organize the dividers and nodes. This pattern allows us to find the full state of a group from the state indicator vector $e_i^{k}$. The state of a group $i$ at any time $t$ is $g_i(t) =e_i^{k}$, which represents that group $i$ is in the $k^{th}$ possible state at time $t$ and full state of that group $i$ can be found from the $k^{th}$ row of the matrix $V_i$. In this model, the joint state of all the groups is needed to understand the network state. The network state or joint state of all the groups at time $t$ is $G(t)$, which is defined as,
 \begin{equation}
     G(t)= g_1(t)\otimes g_2(t) \otimes..........\otimes g_C(t)
 \end{equation}
 Here, $\otimes$ represents the Kronecker product. The dimension of the joint state vector is $\bigg[\binom{\mathcal{N}_1 +M-1}{M-1}\binom{\mathcal{N}_2 +M-1}{M-1}....\binom{\mathcal{N}_C +M-1}{M-1}\bigg] \times 1$. In $G(t)$ all elements are zero except one element corresponding to the network state. The dimension of the joint state vector is less than or equal to the dimension of the individual-based framework. Therefore, $\bigg[\binom{\mathcal{N}_1 +M-1}{M-1}\binom{\mathcal{N}_2 +M-1}{M-1}....\binom{\mathcal{N}_C +M-1}{M-1}\bigg] \leq M^N$. From the network state $G(t)$ at any time $t$, it is possible to infer each group state $g_i(t)$ in the following way,
 \begin{multline}
  g_i(t)=\bigg(1^T_{\binom{\mathcal{N}_1+M-1}{M-1}\times 1}\otimes\cdots\otimes I_{\binom{\mathcal{N}_i+M-1}{M-1}\times \binom{\mathcal{N}_i+M-1}{M-1}}\otimes\\\cdots\otimes1^T_{\binom{\mathcal{N}_C+M-1}{M-1}\times 1}\bigg)G(t)   
 \end{multline}

 
\subsection{Group-level transitions}
All events or transitions are modeled here as independent Poisson processes, therefore the waiting times for events are exponentially distributed. Hence, the system has the memoryless property. An event or transition in a group changes the network state, which is the state transition in the Markov chain. The transitions in the group state are also two types: 1) nodal transition or independent transition, and 2) edge transition or dependent transition. 
\subsubsection{Nodal transition }This transition only depends on the state of a group. It does not depend on the state of its neighboring groups. If a nodal transition from compartment $m$ to compartment $n$ ($m\rightarrow n$) happens in a node of a group $i$ with a rate $\delta$ that means a node in group $i$ moves from compartment $m$ to $n$. Therefore, the group will change its state for this nodal transition from group state $k$ to $l$ with the rate $x_{i,m}\delta$.
\subsubsection{Edge transition and an approximation}
\begin{figure}[htbp]
    \centering
   \includegraphics[width=0.45\textwidth]{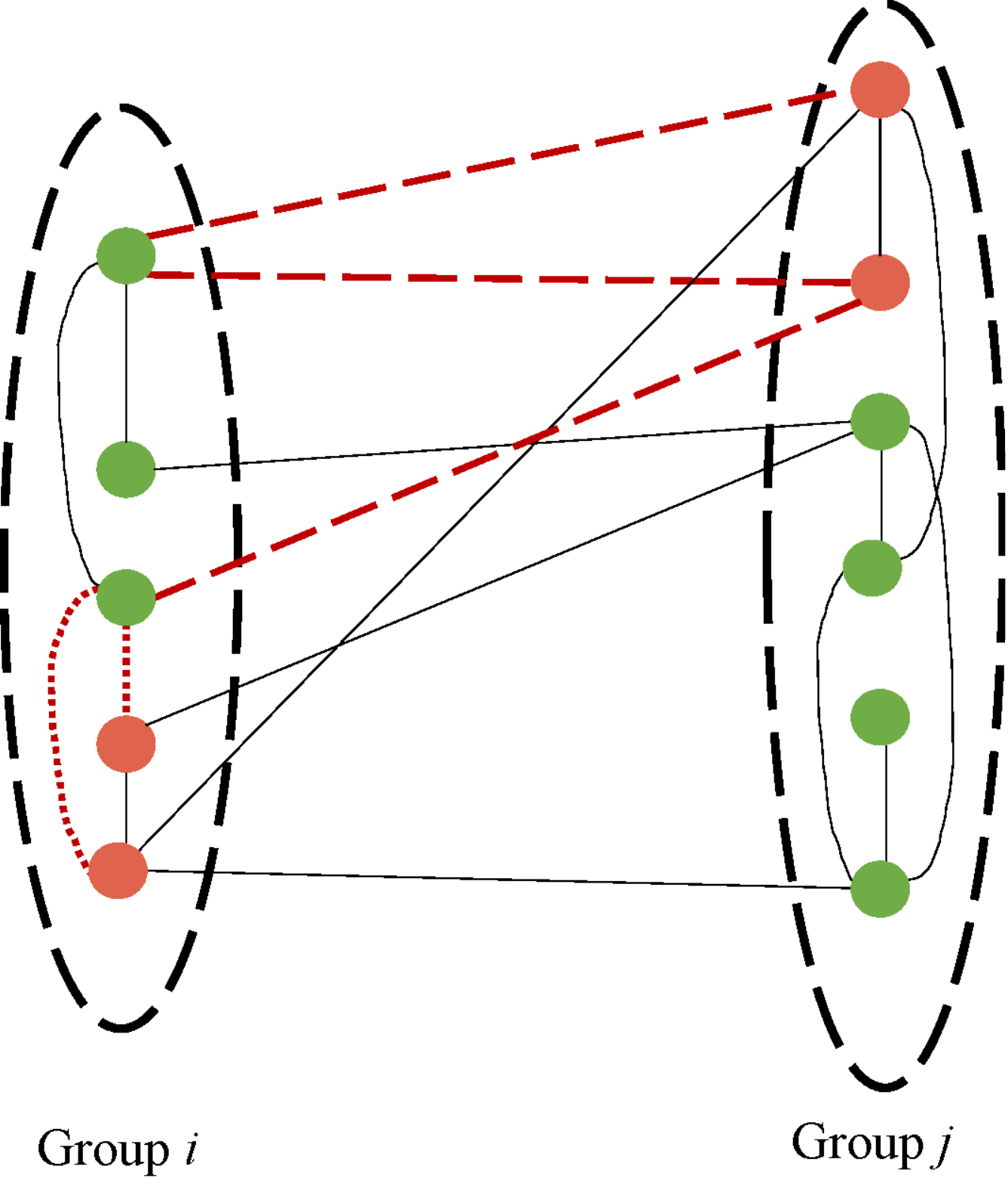}
  \caption{A network with two groups, where nodes are divided into two compartments $m$ and $n$. The green nodes are in the compartment $m$ and the red nodes are in the compartment $n$. For this case, $m$ compartment is the susceptible compartment and $n$ compartment is the infected compartment. The red edges represent the surface of the edge transition $m \rightarrow n$ for the group $i$ (the dashed line edges represent the surface between group $i$ and $j$ and the dotted line edges represent the surface between group $i$ and $i$).}      
\label{surface}
\end{figure}

The edge transition of a group $i$ depends on its own state with the state of its neighboring groups. The edge transition is a dependent transition as it depends on the states of its neighboring groups. When an edge transition happens on a node of a group $i$, it depends on its neighboring nodes, who are distributed in different groups. As an example, the susceptible-to-infected edge transition in group $i$ happens if a susceptible node in group $i$ is in contact with at least an infected node. However, the group-level framework does not contain information about which node is in which compartment. Also, the group-level adjacency matrix cannot tell about the exact neighbors of a node. Therefore, the edge transition in the group-level needs an approximation at the network-level.\\
This approximation is based on the discrete isoperimetric inequality. It is an ancient Greek problem. Isoperimetric inequality focuses on the relationship between the surface and the volume of an object \cite{chung2004discrete, devriendt2017unified}. For this approximation, we need to define the surface and volume in the context of a network. The surface is related to the edges and volume is related to the nodes of a network. An edge is a unit of the surface and a node is a unit of the volume. A nodal transition from compartment $m$ to compartment $n$ depends on the the volume of compartment $m$, that means number of nodes in compartment $m$. An edge transition is more complex than a nodal transition. An edge transition from compartment $m$ to $n$ is proportional to the surface area from compartment $m$ to $n$, which is the number of edges from nodes of compartment $m$ to nodes of compartment $n$. Let, $\mathcal{X}$ is the set of nodes that are in compartment $m$ in group $i$ and $\mathcal{Y}$ is the  set of nodes that are in compartment $n$ in group $j$. Now, $L_{\mathcal{X}\mathcal{Y}}$ is the number of edges, which have one end in a node in $\mathcal{X}$ and another end in a node of $\mathcal{Y}$. The red dashed lines in Fig. \ref{surface} are presenting  $L_{\mathcal{X}\mathcal{Y}}$, where $m$ is the susceptible compartment and $n$ is the infected compartment in the SIS dynamic. The surface for the edge transition $m\rightarrow n$ event between group $i$ and $j$ is,
\begin{equation}
    \text{Surface}_{m\rightarrow n}=  L_{\mathcal{X}\mathcal{Y}}
\end{equation}
\\However, in the group-level approximation, we don't know which node is in which compartment. This model only tells us about the volume of each compartment in each group. Here, we need the approximation which Devriendt et. al. \cite{devriendt2017unified} defined as a topological approximation. This can be defined as,
\begin{equation}
   \text{Surface}_{m\rightarrow n}= L_{\mathcal{X}\mathcal{Y}}\approx \mathcal{A}_g(i,j)|\mathcal{X}||\mathcal{Y}| = Volume
   \label{topoappr}
\end{equation}
Here, $|\mathcal{X}|= x_{i,m}$, $|\mathcal{Y}|= x_{j,n}$, and $d$ is the average node degree and $d\in R$.\\
It is possible to give a bound on the topological approximation from the isoperimetric inequality \cite{chung2004discrete, devriendt2019tighter}. This bound is based on the discrepancy inequality \cite{chung2004discrete},
\begin{equation}
    |L_{\mathcal{X}\mathcal{Y}} - \frac{d}{N}|\mathcal{X}||\mathcal{Y}|| \leq \frac{\theta}{N}\sqrt{|\mathcal{X}|(N- |\mathcal{X}|)|\mathcal{Y}|(N-|\mathcal{Y}|)}
    \label{topoapprobound}
\end{equation}
here, $|d- \sigma_i|\leq \theta$ for $i\neq 0$. $\sigma_i$ are the eigenvalues of the Laplacian matrix of the network for $1\leq i< N$. There is other way to give tighter bound on this approximation derived from the Max Cut problem and the expander mixing lemma \cite{devriendt2019tighter}.\\ For the average node degree, we approximate,
\begin{equation}
    d \approx N\frac{L_{ij}}{\mathcal{N}_i\mathcal{N}_j}
\end{equation}
This approximation comes from a intuition that if nodes of group $i$ has $L_{ij}$ connections with the nodes of group $j$, then nodes of group $i$ has total number of connection in the network is $L_{ij}\frac{N}{\mathcal{N}_j}$. Therefore,
\begin{multline}
     \frac{d}{N}|\mathcal{X}||\mathcal{Y}|= \frac{L_{ij}}{\mathcal{N}_i\mathcal{N}_j} x_{i,m} x_{j,n} = \mathcal{A}_g(i,j)x_{i,m} x_{j,n}
\end{multline}
\subsection{Evolution of the network state }
Evolution of the network state $G(t)$ follows a continuous-time Markov process. Network state is the joint state of all the group states. In the group-based framework, groups are interacting entities, which are jointly Markovian and form a collective system. In the individual-based network model, each node is different with respect to its connections with the neighboring set of nodes, however each node has same set of possible states. In the group-based network model, each group is also different with respect to its connections with the neighboring set of groups, however this model has another complexity that each group can have different sets of possible group states corresponding to their sizes. For example, in an SIS epidemic, each individual node in the individual-based approach has two possible states: susceptible and infected. However, in the group-based approach, each individual group has different number of possible states depends on their size (number of nodes in that group); the number of possible states in a bigger group is larger than a smaller group. Therefore, the group-based Markovian process has another level of complexity than the individual-based one. If the network is in the $G(t)$ state at time $t$, then the evolution of the network state tells us that the network state $G(t+\Delta t)$ after $\Delta t$ time. To obtain $G(t + \Delta t)$, at first we derive the expression for the state of a group $g_i(t+\Delta t)$ at time $(t + \Delta t)$ given that the network is in the $G(t)$ state at time $t$, which is $ Pr[g_i(t+\Delta t)= e^l_i|g_i(t)= e^k_i, G(t)]$. This expression indicates the probability of a transition of a group $i$ from state $k$ to $l$ in a time interval $(t, t+ \Delta t]$. \\
The dynamics of a group state depends on the state of its neighboring groups state. The group state changes when a stochastic event occurs or a node in the group changes its compartment. The waiting time for any event or transition in the network is stochastically independent. We define the transition matrix for a group $i$ as $\theta_i$. The dimension of this matrix is $\binom{\mathcal{N}_i+M-1}{M-1} \times \binom{\mathcal{N}_i+M-1}{M-1}$.\\
Before defining the elements of the $\theta_i$ matrix, we define a new matrix for each type of transition. We name it \textit{transition indication matrix}. For a nodal transition, the \textit{transition indication matrix} will be $\Delta_{ i,\delta_q}$; this matrix indicates a group state transition by a $q$-type nodal transition. In other words, it indicates a nodal transition of a node from compartment $m$ to compartment $n$ with a transition rate $\delta_q$ in the group $i$. The dimension of $\Delta_{\delta_q i}$ is same as the $\theta_i$. If this transition moves the group state from $k$ to $l$, then the elements of $\Delta_{\delta_q i}$ can be defined as,
\begin{equation}
    \Delta_{i,\delta_q}(\mathcal{I},\mathcal{J})= \begin{cases}
     \delta_q&\begin{aligned}
        \text{if $\mathcal{I}=k$ and $\mathcal{J}=l$}\\
        
    \end{aligned}\\
    -\delta_q & \text{if $\mathcal{I}=k$ and $\mathcal{J}= k$}\\
    0 & \text{otherwise}
    \end{cases}
    \label{transitionImatrix}
\end{equation}
Here, $\mathcal{I}$ and $\mathcal{J}$ are index variables for the matrix $\Delta_{i,\delta_q}$. This square matrix indicates all the group-level transitions for the nodal transition $\delta_q$. A simple example of this matrix is given in Eq. \ref{exampletransitionindication} in appendix \ref{Appendixexample}.\\
The definition of the \textit{transition indication matrix} for an edge transition is same as the \textit{transition indication matrix} for a nodal transition, and we denote it as $\Delta_{i,\beta_q}$. The transition  $\beta_q$ represents an edge transition from compartment $m$ to $n$ with rate $\beta_q$. The group transition matrix depends on the node-level transition, as a transition event in the node-level happening at a time $t$ causes the change in the group state.\\
Now, we can define the elements of the transition matrix $\theta_i$, which represents the group transition. If an epidemic model has $q_n$ types of nodal transitions and $q_e$ types of edge transitions, then an element of group transition matrix $\theta_i(k,l)$ has $q_n+q_e$ parts. This element represents the transition from state $k$ to $l$, and each part corresponds to each transition. Even though it has $q_n +q_e$ parts corresponding to $q_n+q_e$ transitions, however only one \textit{transition indication matrix} has a non-zero element in the $(k,l)\textsuperscript{th}$ position in $\theta_i$. A part of $\theta_i(k,l)$ for a $q$ type nodal transition will be,
\begin{equation}
    (e^l_i)^T(\Delta_{i,\delta_q}^T \circ \mathcal{V}_{i,m})e^k_i
\end{equation}
Here, $\circ$ indicates the Hadamard product or element wise product, and each row of $\mathcal{V}_{i,m}$ is equal to the $m$\textsuperscript{th} column of the $V_i$ matrix which is $V_i(:m)$. So, $\mathcal{V}_{i,m}$ is a square matrix with the same dimension of $\Delta_{i,\delta_q}$. Examples of this matrix are given in Eq. \ref{examplevim1} and \ref{examplevim2} in the Appendix \ref{Appendixexample}. 
\\Now, let us consider a $q$-type edge transition from compartment $m$ to $n$ with the rate $\beta_q$, which depends on the $n$ compartmental neighboring nodes. The group state transition rate of a group $i$ for the edge transition $\beta_q$ is ,
\begin{equation}
    \beta_q\sum_{j=1}^{j=C}{u_{i,m}Au_{j,n}}
    \label{beforetopo}
\end{equation}
Here, $u_{i,m}$ is a vector with dimension $1\times N$. Elements correspond to the nodes of the network: $u_{i,m}(n)=1$ if $n$\textsuperscript{th} node is in compartment $m$ and in group $i$, otherwise $u_{i,m}(n)=0$, and  $\sum_{n=1}^{n=N}u_{i,m}(n) = (V_i(:,m))^T g_i(t)$, representing the number of nodes in the $m$ compartment in group $i$ when the group is in $g_i(t)$ state.\\
From the topological approximation we can write the expression (\ref{beforetopo}) as,
\begin{equation}
     \beta_q\sum_{j=1}^{j=C}{X_{i,m}\mathcal{A}_g(i,j)X_{j,n}}
\end{equation}
Here, $X_{i,m}=(V_i(:,m))^T g_i(t)$.\\
Now, we can define the part of $\theta_i(k,l)$ for this edge transition as,
\begin{equation}
(e^l_i)^T\sum_{j=1}^{C}\{\mathcal{A}_g(i,j)X_{j,n}\}  (\Delta_{i,\beta_q}^T \circ \mathcal{V}_{i,m})e^k_i
\end{equation}

Therefore, a element of the group-level transition matrix $\theta_i$ is,
\begin{multline}
\label{qklEquation}
    \theta_i(k,l)= \sum_{ q=1}^{q_n}{ (e^l_i)^T(\Delta_{i,\delta_q}^T\circ \mathcal{V}_{i,m})e^k_i} +\\ \sum_{ q=1}^{q_e}{(e^l_i)^T(\sum_{j=1}^{C}\mathcal{A}_g(i,j)X_{j,n})    (\Delta_{i,\beta_q}^T\circ \mathcal{V}_{i,m})e^k_i}
\end{multline}
Here, $\theta_i(k,l)$ represents the rate for the group transition from state $k$ to $l$. The Eq. (\ref{qklEquation}) has two parts; the first one is for all types of nodal transitions and the second one is for all types of edge transitions. The transition matrix $\theta_i$ is group specific; different groups will have different $\theta_i$ matrix.  \\
The transition from group state $k$ to $l$ is an independent Poisson process, which occurs in $(t,t+ \Delta t]$ time interval. Therefore,
\begin{equation}
\label{probabilityEquation}
    Pr[g_i(t+\Delta t)= e^l_i|g_i(t)= e^k_i, G(t)]= \theta_i(k,l)\Delta t + o(\Delta t)
\end{equation}

Eq. (\ref{probabilityEquation}) will be used to derive the time evolution of the network state. The group transition matrix $\theta_i$ give us the description of the group-level transition. The evolution of the network state $G(t)$ is a continuous-time Markov process, where actual Markov states are the possible network states. This process is fully characterized by a systems of differential equations named as the Kolmogorov differential equations for a given initial condition. The procedure to derive the Kolmogorov differential equations for a Markov chain from the one state transition rates is a standard process, which is described in \cite{ross1996stochastic, van2014performance}. \\
The expected value of a group state in the next time step, when the network is in the $G(t)$ state, can be obtained from Eq. (\ref{qklEquation}) and (\ref{probabilityEquation}),
\begin{multline}
\label{onesideEquation}
    E[g_i(t+\Delta t)|G(t)]= \sum_{ q=1}^{q_n}{ (\Delta_{i,\delta_q}^T\circ \mathcal{V}_{i,m})g_i(t)}\Delta t \\+ \sum_{ q=1}^{q_e}{(\sum_{j=1}^{C}\mathcal{A}_g(i,j)X_{j,n}) (\Delta_{i,\beta_q}^T\circ \mathcal{V}_{i,m})g_i(t)}\Delta t + g_i(t) + o(\Delta t)
\end{multline}
Now, considering the expected value of both sides in Eq. (\ref{onesideEquation}),
\begin{multline}
\label{twosideEquation}
  E[ E[g_i(t+\Delta t)|G(t)]]=E[g_i(t+\Delta t)] = \\\sum_{ q=1}^{q_n}{ (\Delta_{i,\delta_q}^T\circ \mathcal{V}_{i,m})E[g_i(t)]}\Delta t \\+ \sum_{ q=1}^{q_e}{  (\Delta_{i,\beta_q}^T\circ \mathcal{V}_{i,m})E[h_i(t)g_i(t)]}\Delta t + E[g_i(t)] + E[o(\Delta t)]
\end{multline}
In Eq. (\ref{twosideEquation}), the expression for $h_i(t)$ is,
\begin{equation}
    h_i(t)=\sum_{j=1}^{C}\mathcal{A}_g(i,j)X_{j,n}
\end{equation}
Here, $h_i(t)$ is the number of neighbors of group $i$, who are in the compartment $n$. And compartment $n$ is the influencer compartment for the $q$\textsuperscript{th} edge transition.  \\
After rearranging the Eq. (\ref{twosideEquation}) as follows,
\begin{multline}
    \frac{E[g_i(t+\Delta t)]- E[g_i(t)]}{\Delta t}= \sum_{ q=1}^{q_n}{ (\Delta_{i,\delta_q}^T\circ \mathcal{V}_{i,m})E[g_i(t)]} \\+ \sum_{ q=1}^{q_e}{   (\Delta_{i,\beta_q}^T\circ \mathcal{V}_{i,m})^TE[h_i(t)g_i(t)]}  +\frac{E[o(\Delta t)]}{\Delta t} 
    \label{differencee}
\end{multline}
Let $\Delta t \rightarrow 0$, so the Eq. (\ref{differencee}) will become,
\begin{multline}
  \frac{d}{dt}E[g_i(t)]=   \sum_{ q=1}^{q_n}{ (\Delta_{i,\delta_q}^T\circ \mathcal{V}_{i,m})E[g_i(t)]} \\+ \sum_{ q=1}^{q_e}{   (\Delta_{i,\beta_q}^T\circ \mathcal{V}_{i,m})^TE[h_i(t)g_i(t)]}
  \label{differetiale}
\end{multline}
The differential equation of the joint state or the network state can be written as,
\begin{equation}
    \frac{d}{dt}E[G]= \Theta E[G]
    \label{exacte}
\end{equation}
This is the time evolution of the network state. Here, $\Theta$ is the network transition matrix of the underlying Markov process, this closed set of differential equation can fully characterize the network state. The rate of change between network states is described by the $\Theta$ matrix, which is equivalent to the infinitesimal $Q$ matrix of the individual-based approach \cite{van2009virus}. Any element in the $\Theta$ matrix $\Theta_{\mathcal{K} \mathcal{L}}$ represents the rate of change from network state $\mathcal{K}$ to $\mathcal{L}$. The dimension of the $\Theta$ is $\bigg[\binom{\mathcal{N}_1 +M-1}{M-1}\binom{\mathcal{N}_2 +M-1}{M-1}....\binom{\mathcal{N}_C +M-1}{M-1} \bigg]\times \bigg[\binom{\mathcal{N}_1 +M-1}{M-1}\binom{\mathcal{N}_2 +M-1}{M-1}....\binom{\mathcal{N}_C +M-1}{M-1}\bigg] $. The derivation of the network transition matrix $\Theta$ for group-based framework are given in Appendix \ref{appendixEMP} and we also provide a simple example of an SIS spreading process in a network with two groups in Appendix \ref{Appendixexample}. The network state evolution in the group-based structure is a multidimensional birth-death process, which is a special case of the continuous-time Markov process. The Eq. (\ref{exacte}) fully describes the dynamics of the underlying system, which is the Kolmogorov differential equation of the Markov process with $\bigg[\binom{\mathcal{N}_1 +M-1}{M-1}\binom{\mathcal{N}_2 +M-1}{M-1}....\binom{\mathcal{N}_C +M-1}{M-1} \bigg]$ states. This is not an exact Markov process, this is a approximated Markov process because of the topological approximation. A strong bound on this approximation can be given from the discrete isoperimetric inequality (Eq. \ref{topoapprobound}).
\section{ Mean-field Approximation of the GgroupEM }
\label{UMFF}
\subsection{Mean-field equations}
\begin{figure}[htb]
    \centering
    \subfloat[]{\includegraphics[width=1.73in]{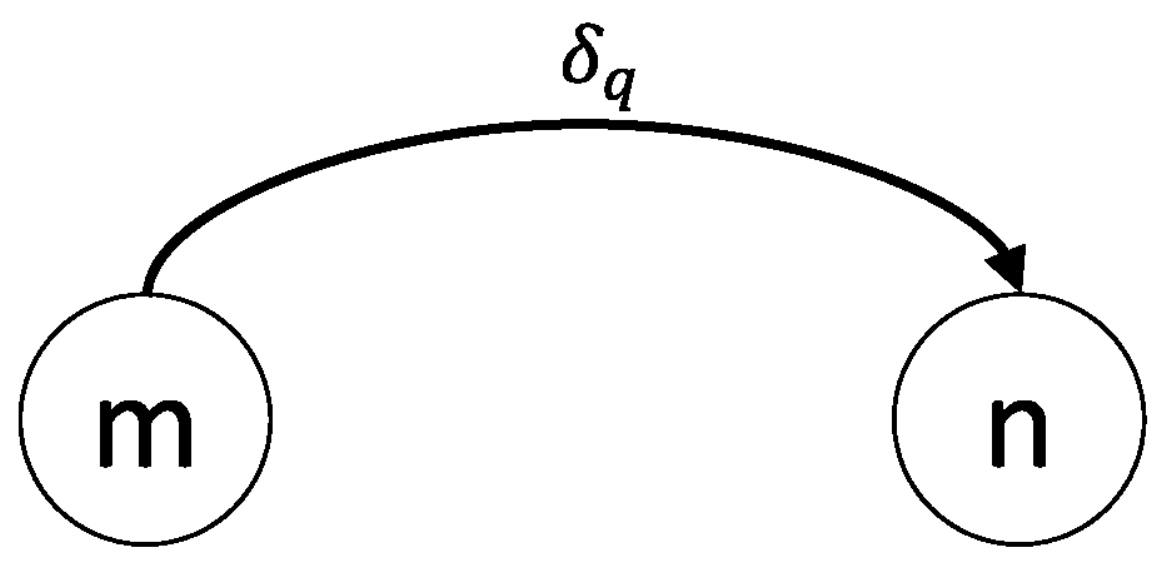}\label{nodeleveltr}}
    
    \subfloat[]{\includegraphics[width=3.63in]{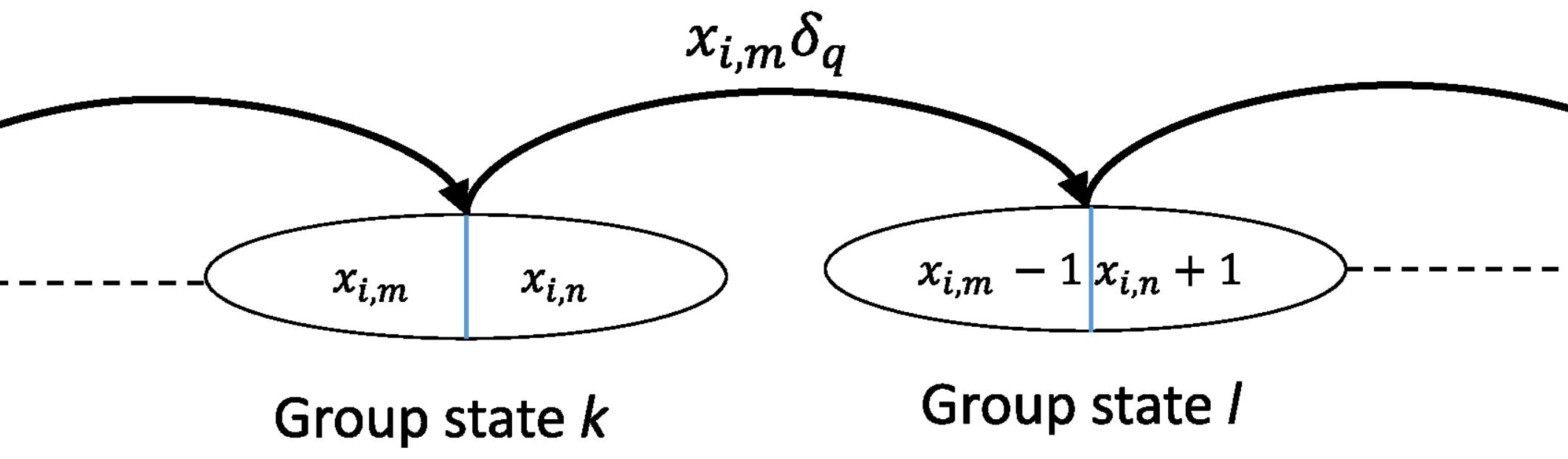}\label{groupleveltr}}
       
  \caption{(a) Node-level transition: nodal transition of the state of a node with $\delta_q$ rate, (b) group-level transition: transition of the state of a group $i$ for this nodal transition. }      

\label{transition}
\end{figure}

The Eq. (\ref{differetiale}) and (\ref{exacte}) contains higher order moment terms $E[h_i(t)g_i(t)]$. This framework has assumed that states of individual groups are independent random variables and invoke moment-closure approximation for those higher order moment. This approximation allows to assume the covariance between two random variable $h_i(t)$ and $g_i(t)$ is zero.  
From the first moment-closure approximation we can write,
\begin{equation}
    Cov[h_i(t)g_i(t)]\approx 0
\end{equation}
\begin{multline}
  \Rightarrow  E[h_i(t)g_i(t)]\approx E[h_i(t)]E[g_i(t)]\\\approx\sum_{j=1}^{C}\mathcal{A}_g(i,j)(V_j(:,n))^TE[ g_j(t)]E[g_i(t)]
\end{multline}
The mean-field equation of the group-based framework will be,
\begin{multline}
  \frac{d}{dt}E[g_i(t)]= \sum_{ q=1}^{q_n}{ (\Delta_{i,\delta_q}^T\circ \mathcal{V}_{i,m})E[g_i(t)]} +\\ \sum_{ q=1}^{q_e}{\Bigg(\sum_{j=1}^{C}\mathcal{A}_g(i,j)(V^{g}_j(:,n))^TE[ g_j(t)] \Bigg)  (\Delta_{i,\beta_q}^T\circ \mathcal{V}_{i,m})^TE[g_i(t)]}
  \end{multline}
This is a first-order closure approximation leading to a unified mean field equation. The higher order equation is given in \cite{devriendt2017unified}. \\
Now, we will provide the equations in the compartmental level for a group. These compartmental equations allow us to describe the evolution of the expected values through the $(M-1)C$ ordinary differential equations. If a node in a group move its compartment from $m$ to $n$ with a rate $\delta_q$, then the group state will change with rate $x_{i,m}\delta_q$. Fig. \ref{nodeleveltr} is presenting that a node in group $i$ is changing its state from compartment $m$ to $n$ through a $q$-type nodal transition and Fig \ref{groupleveltr} is presenting that the group state has changed from state $k$ to $l$ for the nodal transition in Fig. \ref{nodeleveltr}. So, the population for the $m$ compartment will be, 
\begin{equation}
    \frac{d}{dt}E[x_{i,m}]= -\delta_q E[x_{i,m}]
\end{equation}
and population for the $n$ compartment will be,
\begin{equation}
    \frac{d}{dt}E[x_{i,n}]= \delta_q E[x_{i,m}]
\end{equation}
So, 
\begin{equation}
    \frac{d}{dt}E[X_i]= Q_{\delta_q}^T E[X_i]
\end{equation}
The transition matrix $Q_{\delta_q}$ represents a nodal transition from compartment $m$ to $n$ with rate $\delta_q$. It has a dimension $M\times M$, where $Q_{\delta_q}(m,m)= -\delta_q$, $Q_{\delta_q}(m,n)= \delta_q$ and zero otherwise. This matrix has the form of a Laplacian matrix.\\

The general equation for all type of transitions is,
\begin{multline}
\label{mf equation}
    \frac{d}{dt}E[X_i]=\sum_{ q=1}^{q_n}Q_{\delta_q}^TE[X_i] \\+\sum_{q=1}^{q_e}\bigg( \sum_{j=1}^C \mathcal{A}_g(i,j) E[X_{j,n}]\bigg)Q_{\beta_q}^TE[X_i]
\end{multline}

If the fraction of node in each compartment in a group $i$ is $\rho_i= \frac{E[X_i]}{\mathcal{N}_i}$, where $\rho_i= [\rho_{i,1}, \rho_{i,2}.....,\rho_{i,M}]^T$ and  $\sum_{m=1}^M{\rho_{i,m}}=1$ then the Eq. (\ref{mf equation}) can be written as,

\begin{multline}
   \dot{\rho_i} =\sum_{ q=1}^{q_n}Q_{\delta_q}^T\rho_i +\sum_{q=1}^{q_e}\bigg( \sum_{j=1}^C  \frac{L_{ij}}{\mathcal{N}_i} \rho_j(n)\bigg)Q_{\beta_q}^T\rho_i
   \label{mfe1}
\end{multline}

Summarizing, GgroupEM framework has two approximations: 1) topological approximation and 2) moment-closure approximation. We only know about error bound of the topological approximation from isoperimetric inequality. However, the error bound for the moment-closure approximation of this framework is not known \cite{devriendt2017unified}. \\
Group-based mean-field equations for some epidemic models are given below.
\subsection{Examples}
\subsubsection{Susceptible-infected-susceptible}
\begin{figure}[htbp!]
     \centering
    \subfloat[]{\includegraphics[width=1.73in]{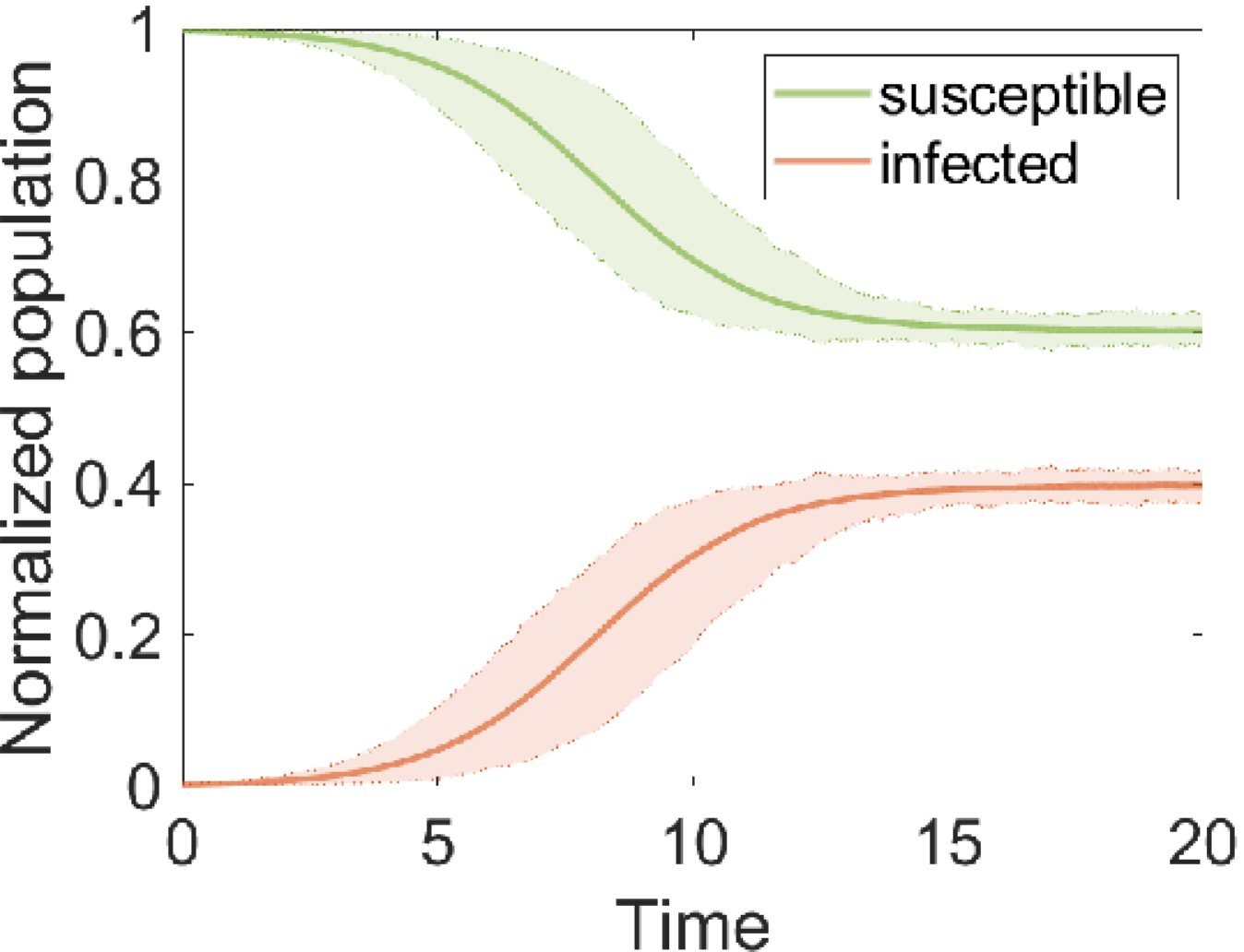}}
    \hfill
    \subfloat[]{\includegraphics[width=1.73in]{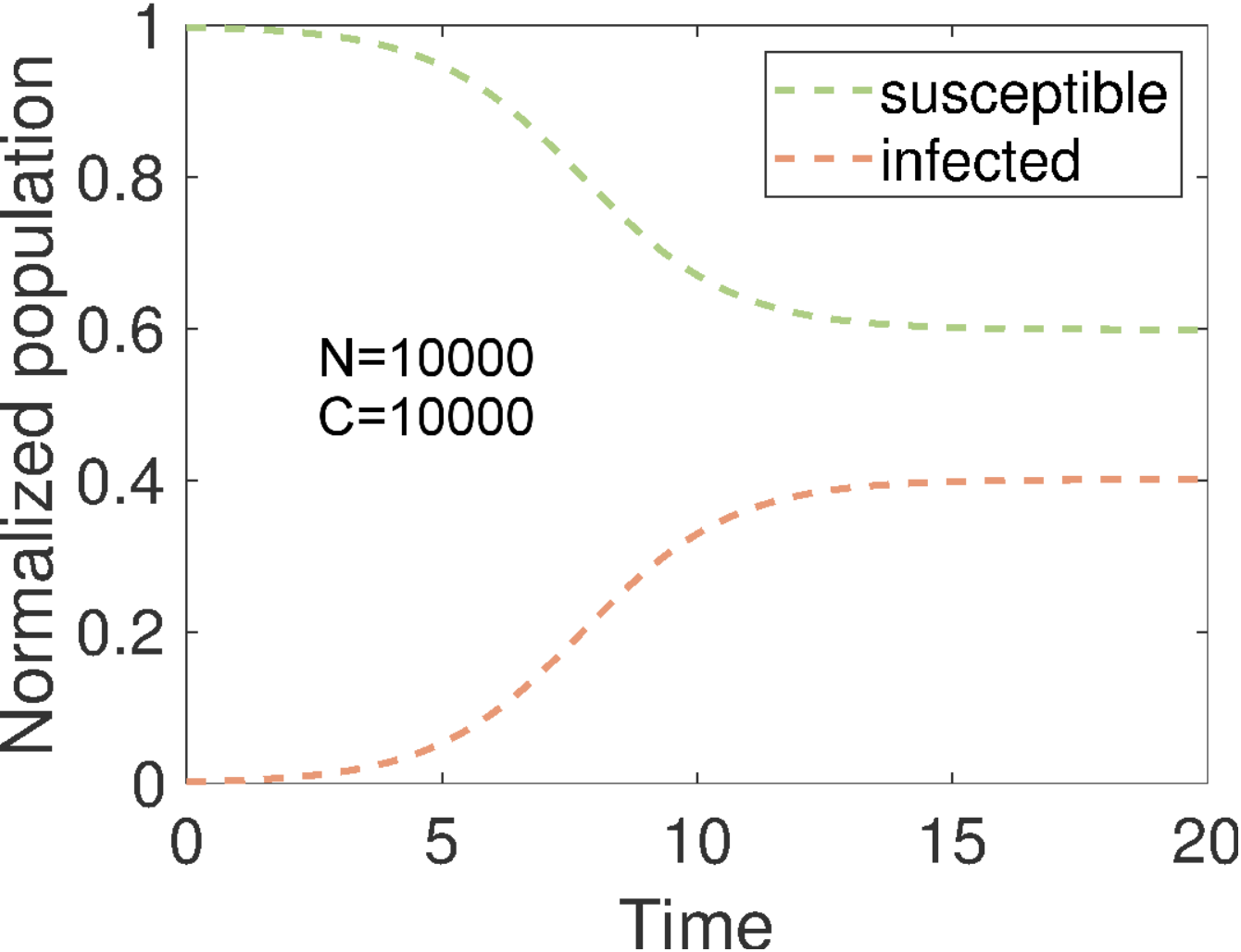}}
   
    \subfloat[]{\includegraphics[width=1.73in]{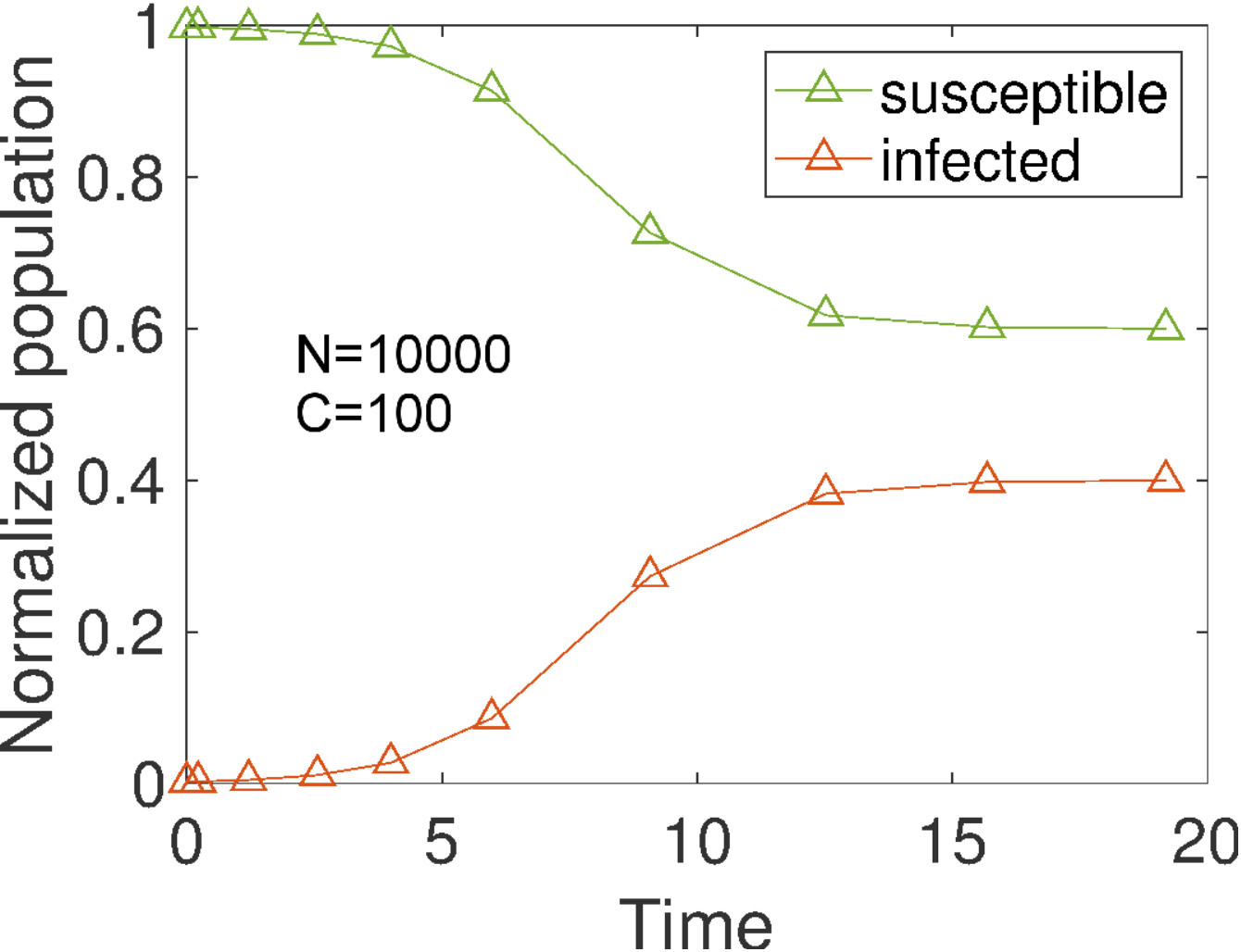}}
      \hfill
    \subfloat[]{\includegraphics[width=1.73in]{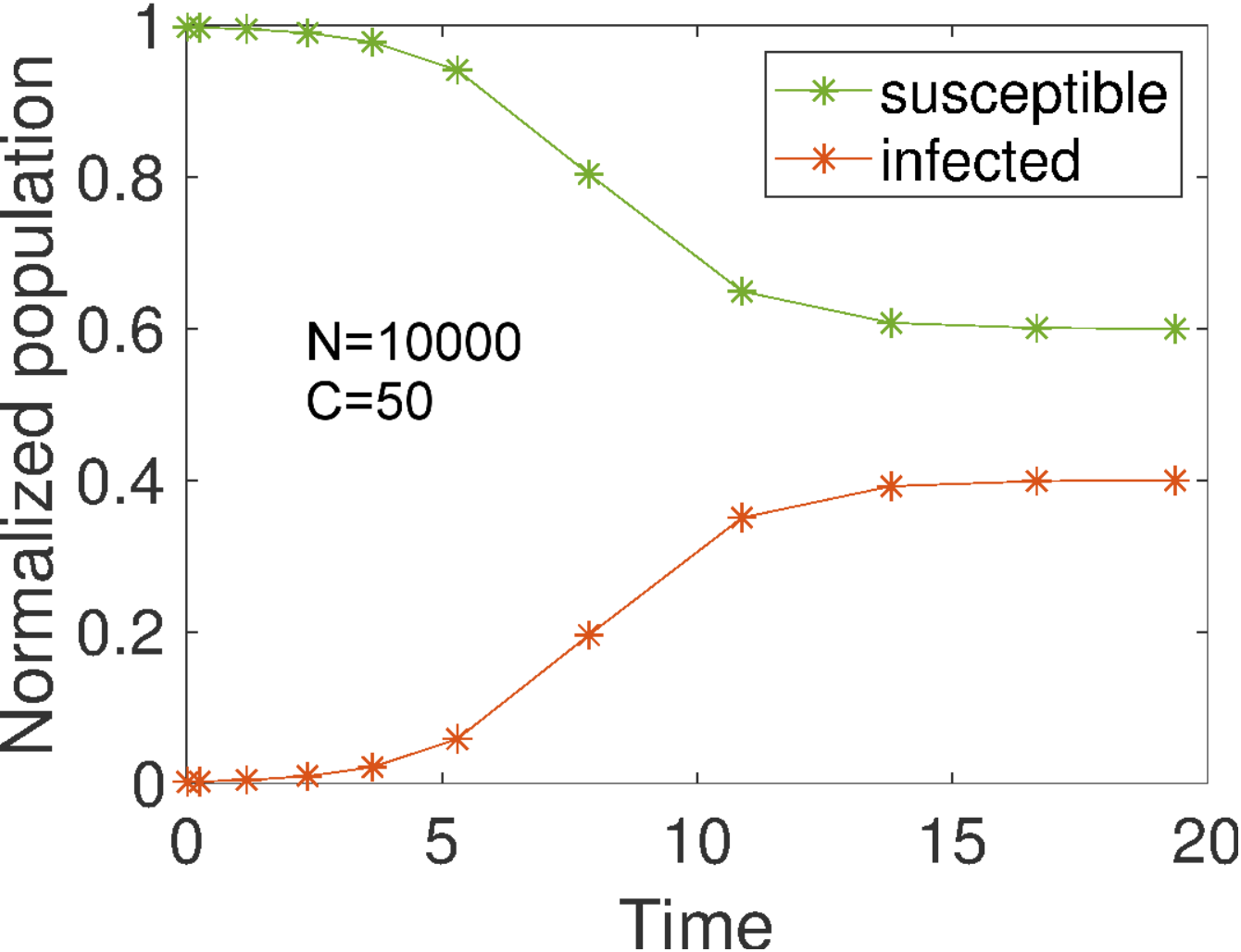}}
    
    \subfloat[]{\includegraphics[width=1.73in]{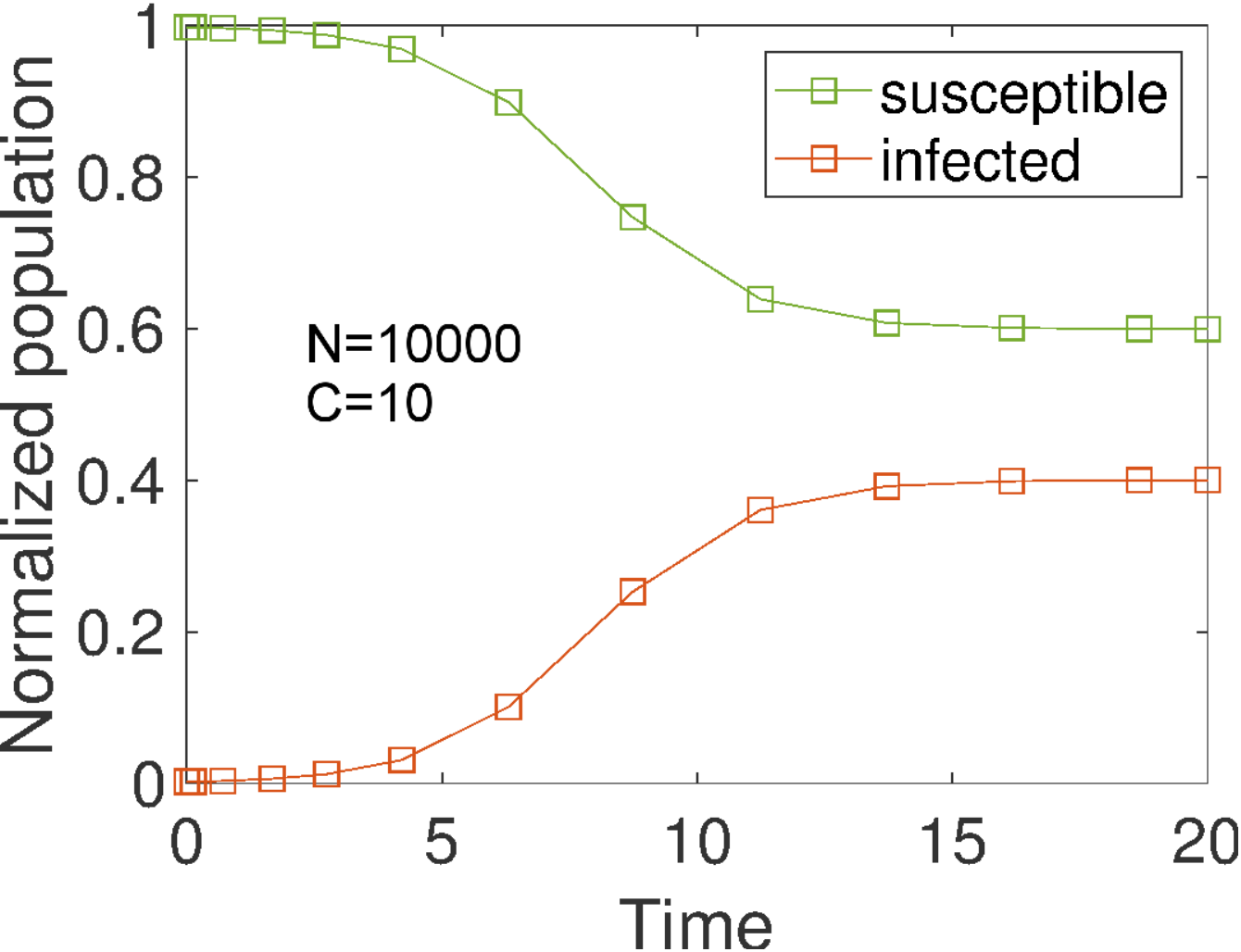}}
      \hfill
    \subfloat[]{\includegraphics[width=1.73in]{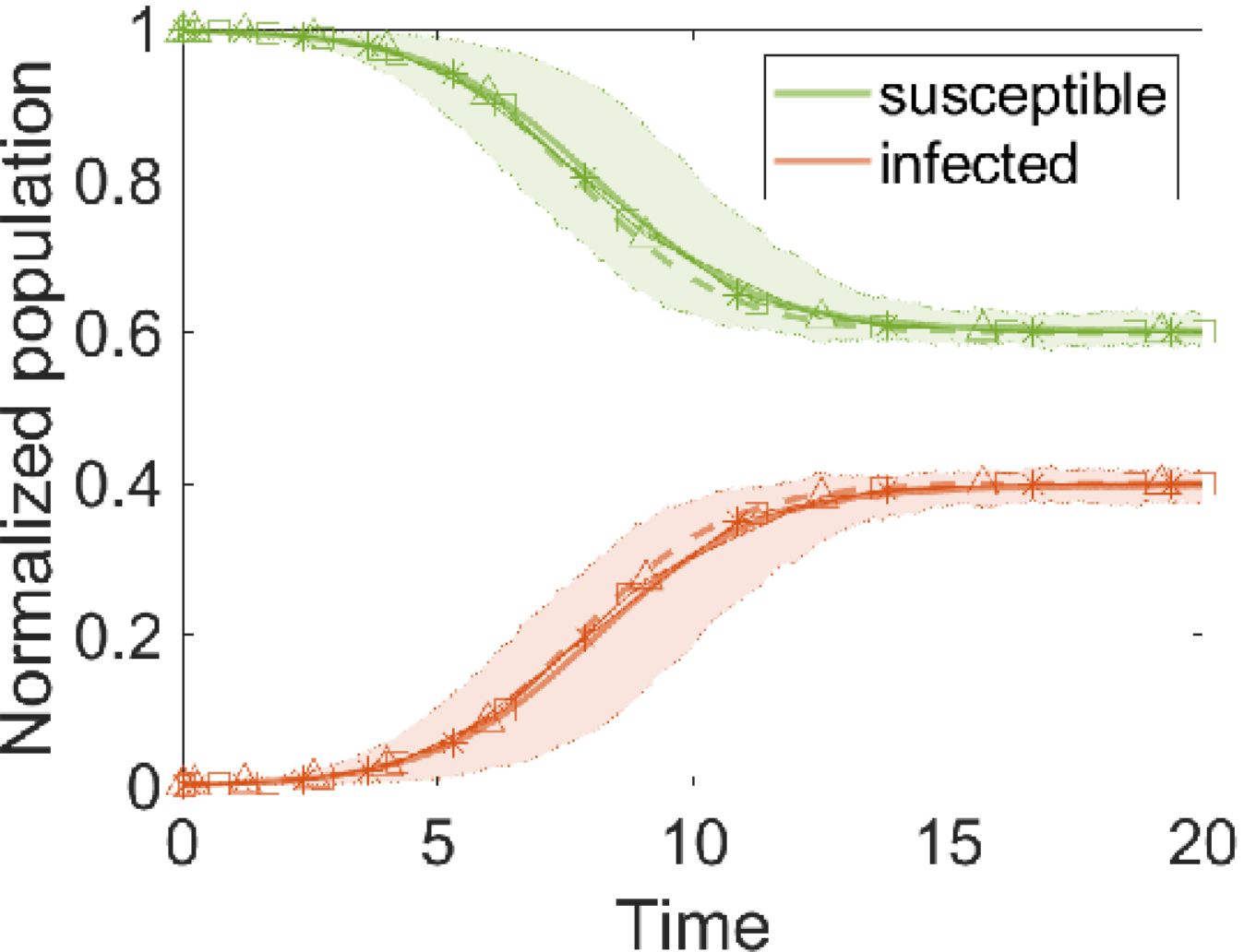}}
  \caption{Results for an SIS epidemic in the Erdos-Renyi random network $(N= 10000, p=0.01)$; a) Stochastic numerical simulation of the Exact Markov process of the individual based approach, solid lines represent the average of the 200 simulations and shaded areas represent region of the stochastic simulation, b) Individual-based: $N=C= 10000 , \mathcal{N}_1= \mathcal{N}_2=.....=\mathcal{N}_C=1$, simulation time $=10.409s$, c) group-based: $C= 100 =1\%N ,  \mathcal{N}_1= \mathcal{N}_2=.....=\mathcal{N}_C=100$, simulation time $=0.084s $, d) group-based: $C= 50 =0.5\%N,  \mathcal{N}_1= \mathcal{N}_2=.....=\mathcal{N}_C=200$, simulation time $=0.049s $, e) group-based: $C= 10 =0.1\%N,  \mathcal{N}_1= \mathcal{N}_2=.....=\mathcal{N}_C=1000$, simulation time $=0.016s $, f) merging of all sub-plots a-e. }      
\label{SISER}
\end{figure}

The SIS model \cite{van2009virus} has two types of transitions; one is an edge transition (susceptible to infected) and the other is a nodal transition (infected to susceptible). The infected compartment is the influencer compartment for the edge transition. The mean-field equation for the group-based framework of SIS epidemic model can be written as,
\begin{multline}
\label{SISExpected}
 \begin{bmatrix}
 \dot{\rho_{i,S}}\\ \dot{\rho_{i,I}}
 \end{bmatrix}=  \bigg( \sum_{j=1}^C  \frac{L_{ij}}{\mathcal{N}_i} \rho_{j,I}\bigg)\underbrace{\begin{bmatrix} 
 -\beta & \beta\\ 0 & 0
 \end{bmatrix}^T}_{Q_\beta^T \text{ matrix}} \begin{bmatrix} \rho_{i,S}\\ \rho_{i,I}\end{bmatrix}\\+\underbrace{\begin{bmatrix} 
 0 & 0\\ \delta & -\delta
 \end{bmatrix}^T}_{Q_\delta^T \text{ matrix}} \begin{bmatrix} \rho_{i,S}\\ \rho_{i,I}\end{bmatrix}
\end{multline}
Here, $\rho_{i,S}$ and $\rho_{i,I}$ represent the fraction of susceptible and infected nodes in the group $i$ and  $\rho_{i,S} + \rho_{i,I} =1$ at any time $t$. The first part and second part in the Eq. (\ref{SISExpected}) is for the edge transition $S\rightarrow I$ (susceptible to infected) and nodal transition $I\rightarrow S$ (infected to susceptible) respectively. The rate for edge transition is $\beta$ and the rate for nodal transtion is $\delta$. This process has $(2-1)C$ ordinary differential equations. \\
The dynamics of an SIS epidemic model for an Erd{\"o}s-R{\'e}nyi $(N,p)$ random network \cite{erdos1959random} is given in Fig. \ref{SISER}. This random network has $N= 10000$ nodes and $p= 0.01$ connection probability. The number of edges and average node degree of this network are $999903$ and $200$ respectively.
The normalized population in different compartments of stochastic numerical simulation of the individual-based approach for SIS is presenting in Fig \ref{SISER}a. The mean-field approximation for the individual-based is given in  \ref{SISER}b. The  Group-based approachs are presenting in Fig \ref{SISER}c-e. For each case, $\beta = 0.0167$ and $\delta = 1$. As a initial condition, we have started the epidemic from $0.2\%$ infected nodes. An summary of the results is given in Table \ref{table_mfa}.

\subsubsection{Susceptible-infected-recovered}
\begin{figure}[h!]
    \centering
   \centering
    \subfloat[]{\includegraphics[width=1.73in]{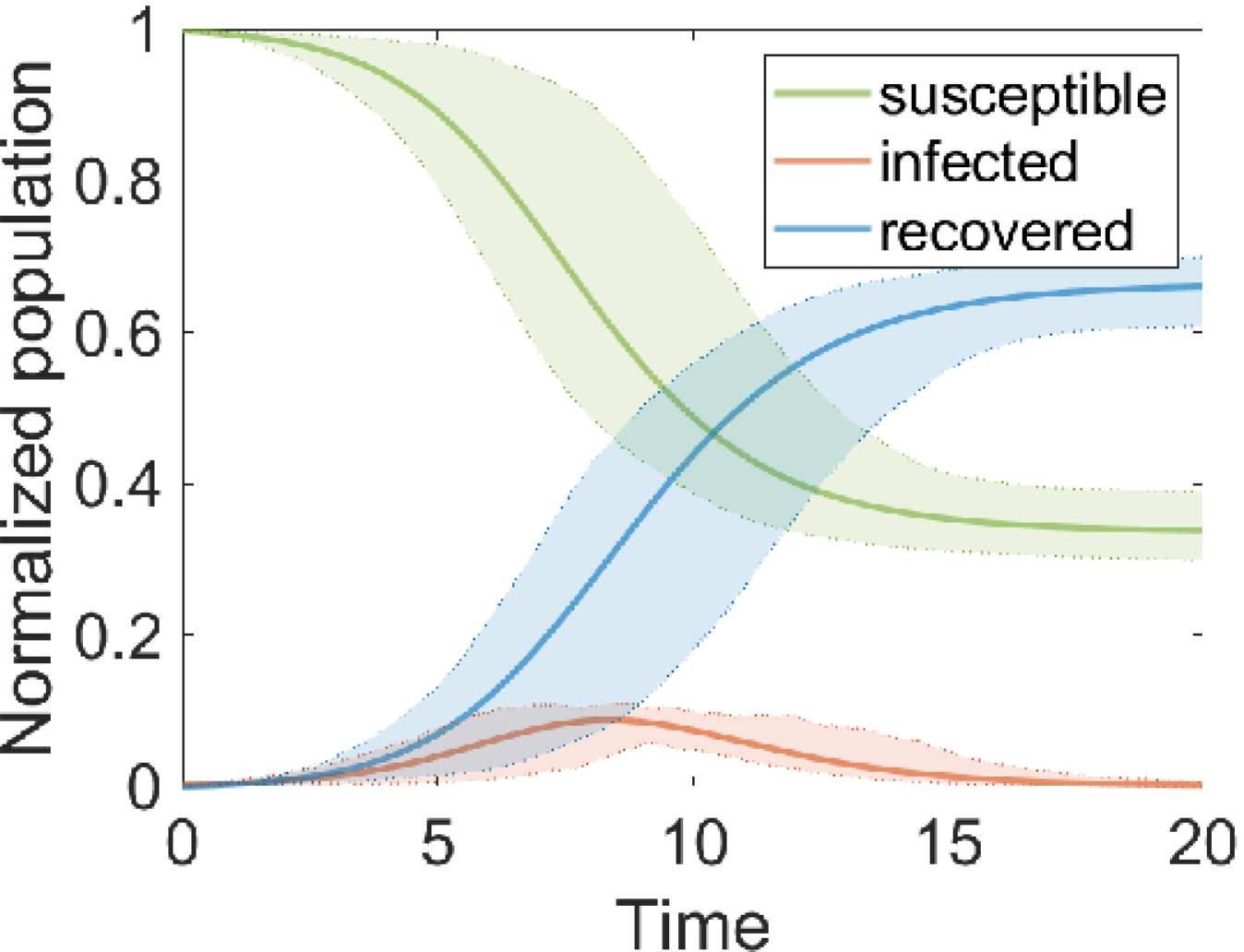}}
    \hfill
    \subfloat[]{\includegraphics[width=1.73in]{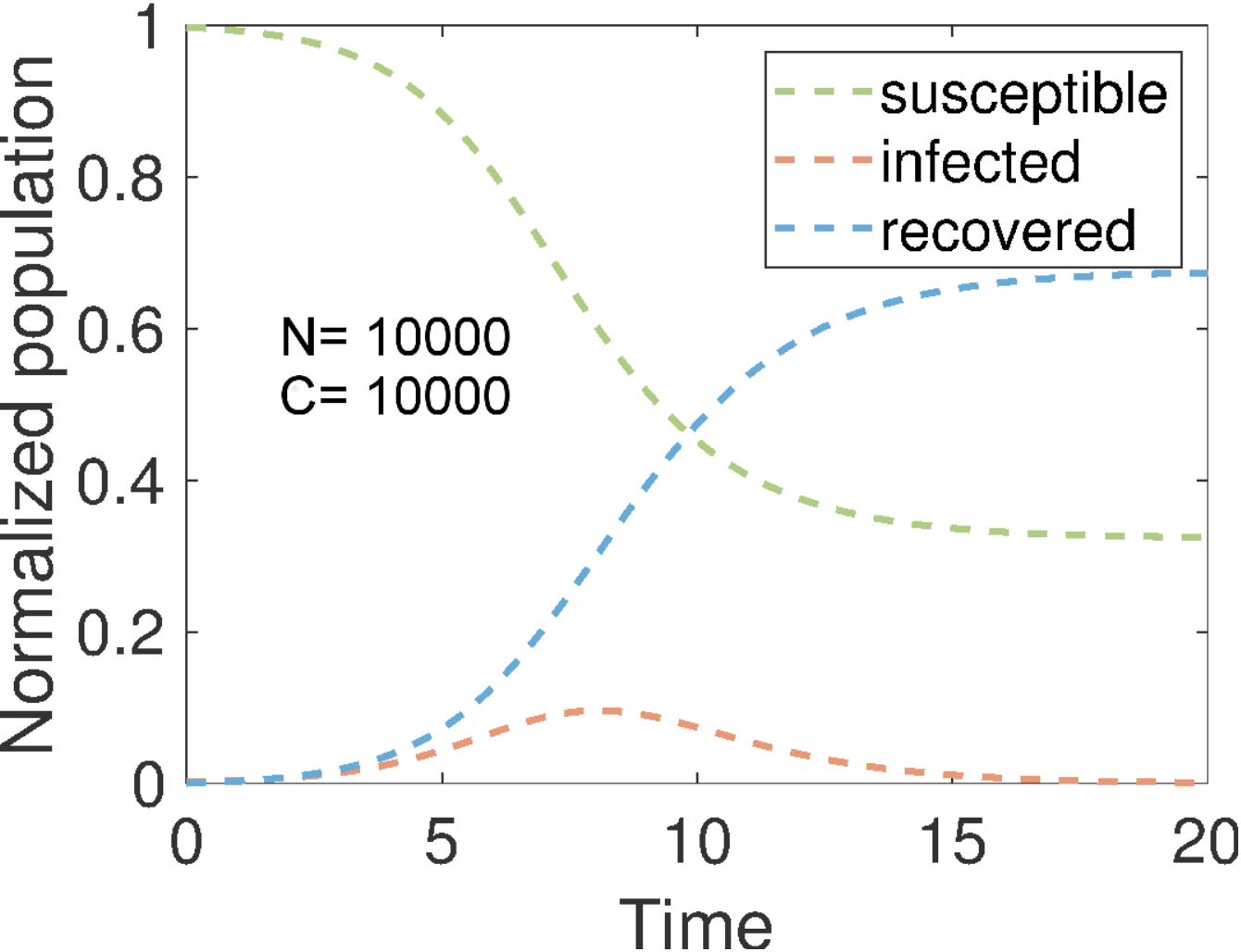}}
   
    \subfloat[]{\includegraphics[width=1.73in]{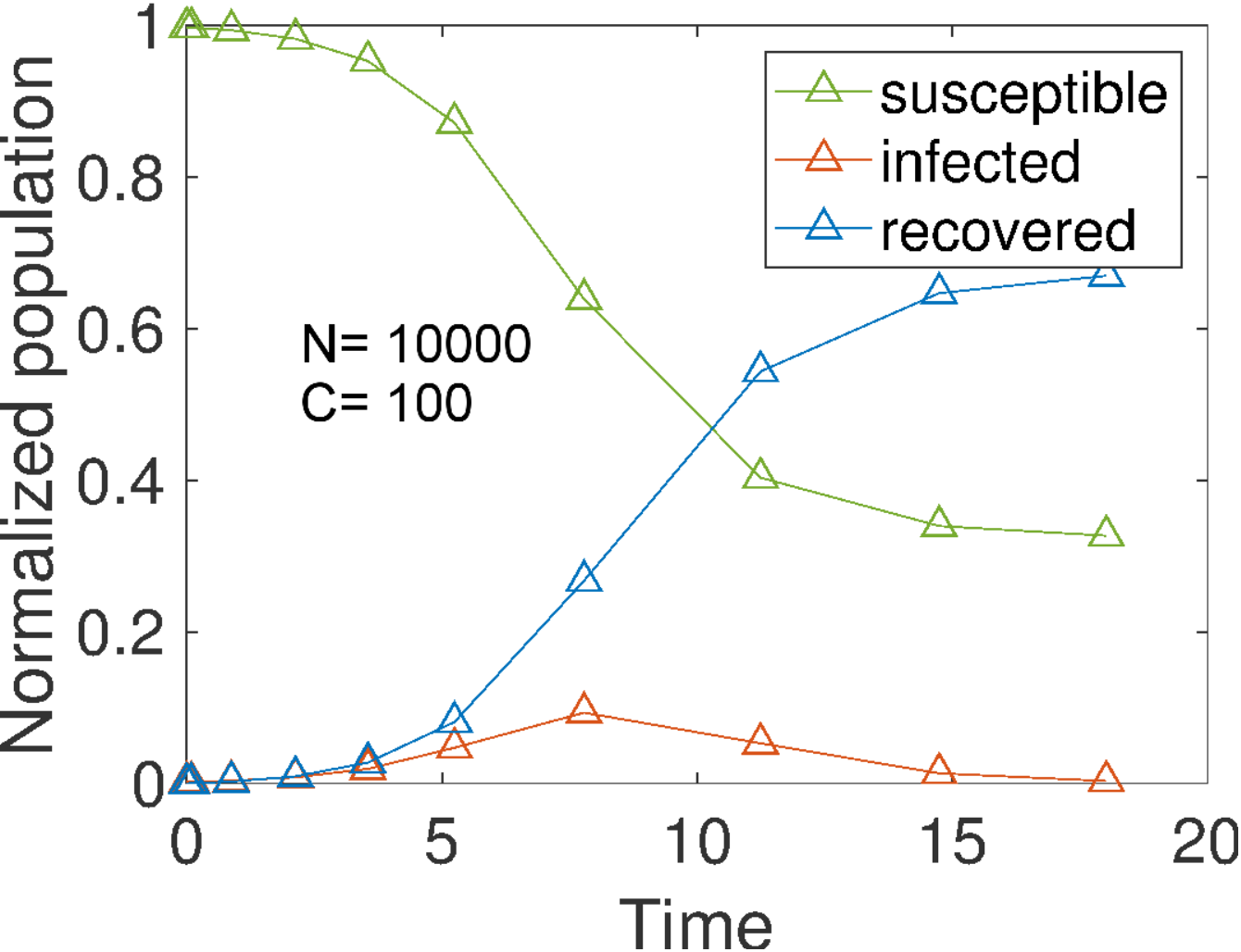}}
      \hfill
    \subfloat[]{\includegraphics[width=1.73in]{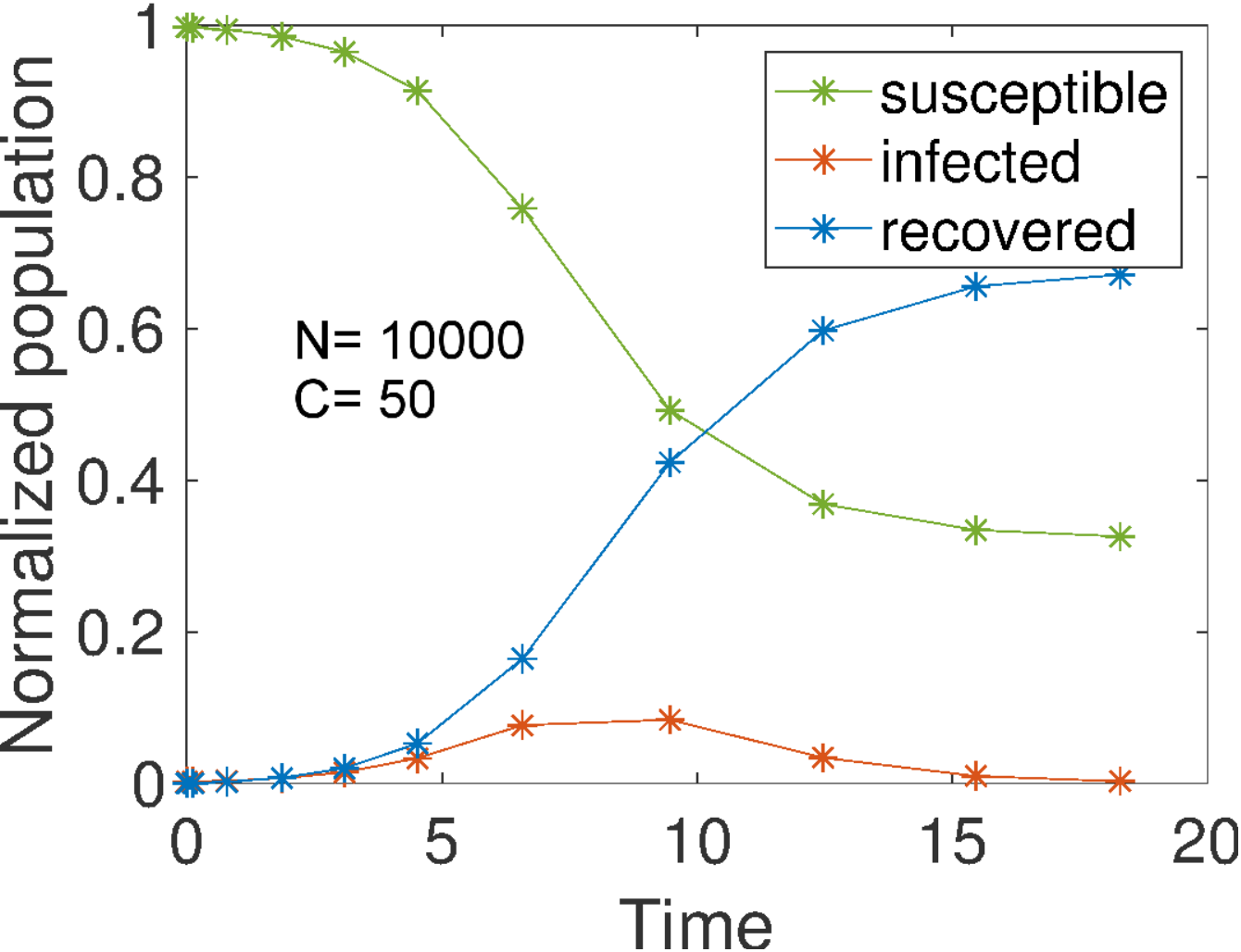}}
    
    \subfloat[]{\includegraphics[width=1.73in]{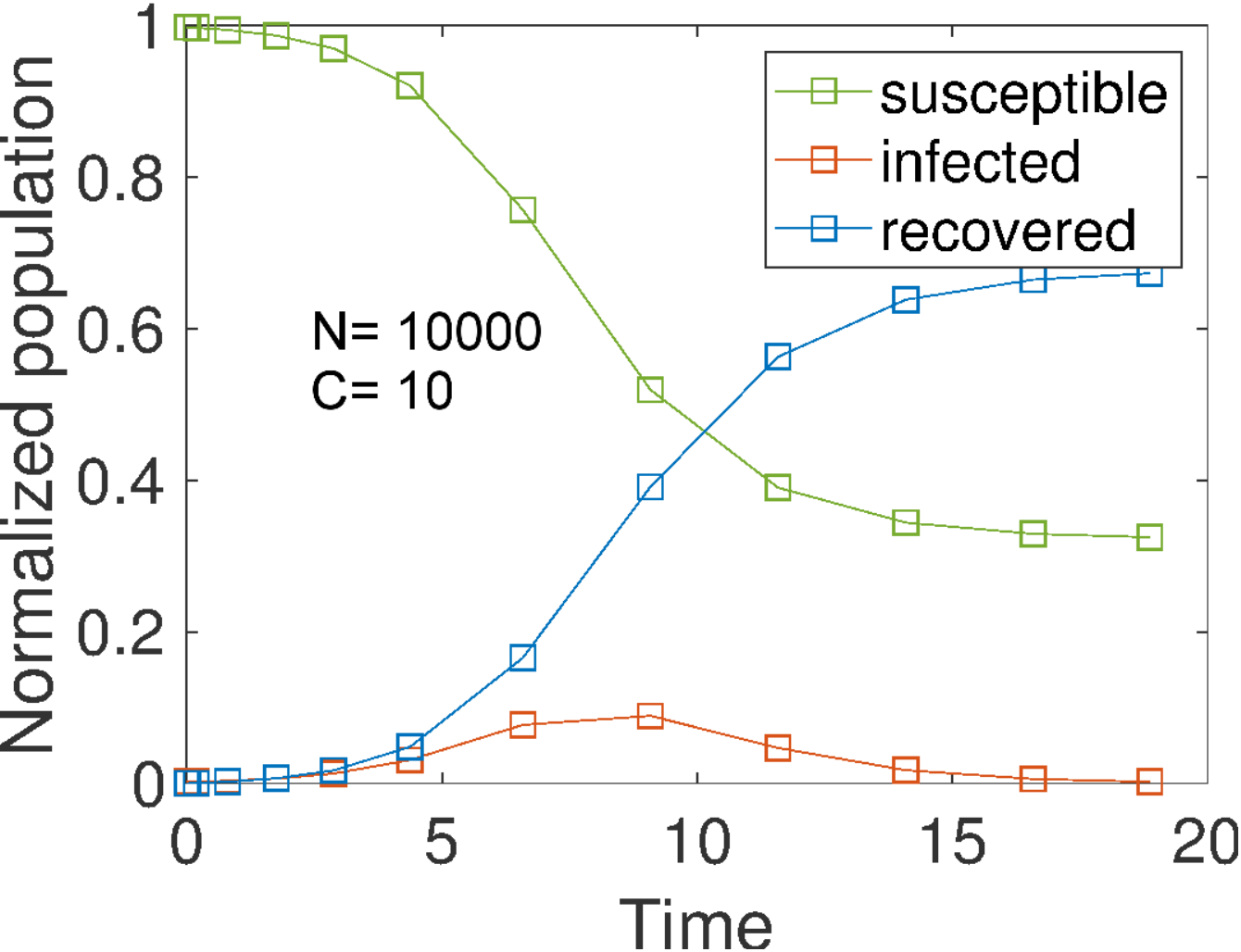}}
      \hfill
    \subfloat[]{\includegraphics[width=1.73in]{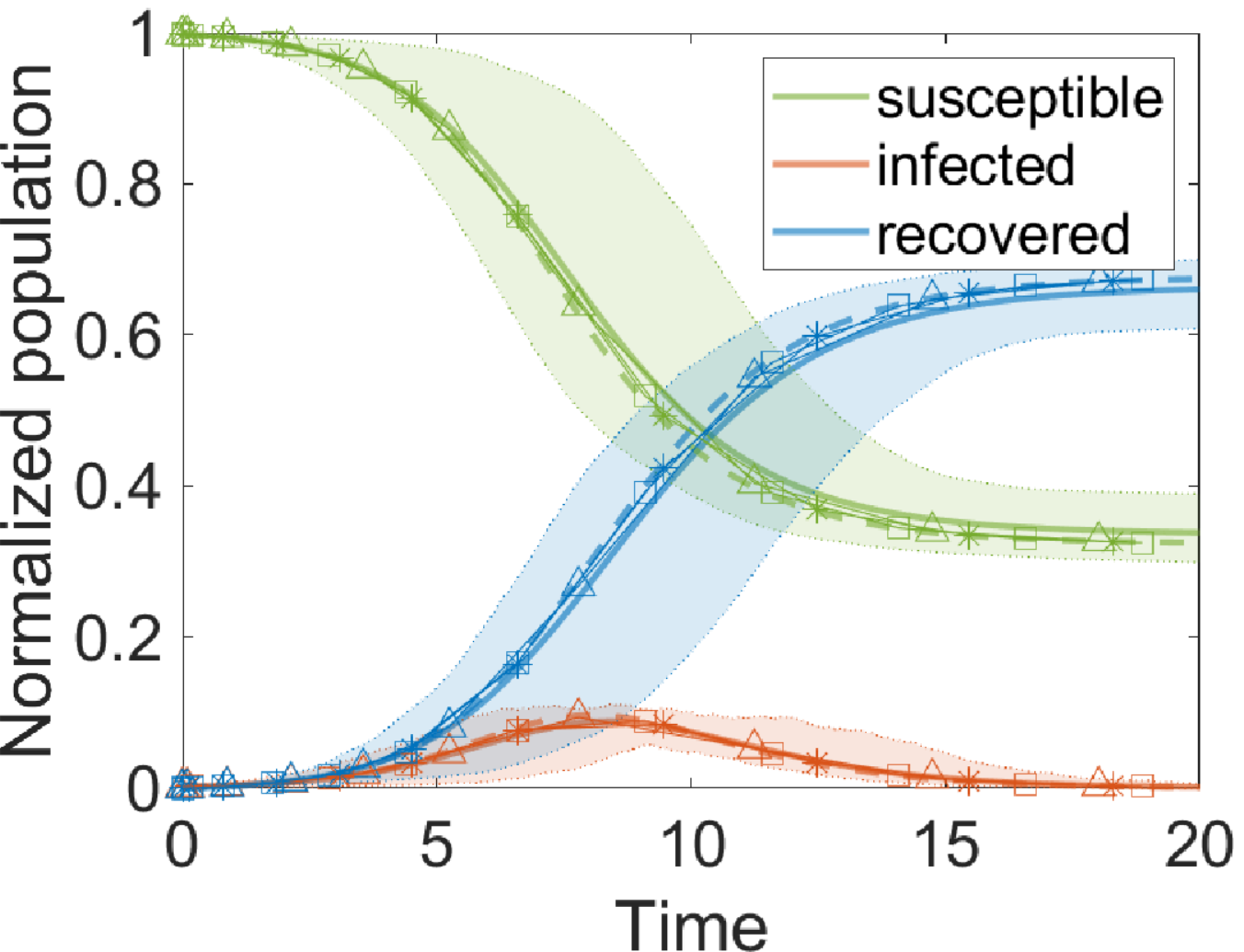}}
       
  \caption{Results for an SIR epidemic in the Erd{\"o}s-R{\'e}nyi network $(N= 10000, p=0.01)$; a) Stochastic numerical simulation of the Exact Markov process of the individual based approach, solid lines represent the average of the 200 simulations and shaded areas represent region of the stochastic simulation, b) individual-based: $N=C= 10000 , \mathcal{N}_1= \mathcal{N}_2=.....=\mathcal{N}_C=1$, simulation time $=12.183s $, c) group-based: $C= 100 =1\%N ,  \mathcal{N}_1= \mathcal{N}_2=.....=\mathcal{N}_C=100$, simulation time $=0.0.088s $, d) group-based: $C= 50 =0.5\%N,  \mathcal{N}_1= \mathcal{N}_2=.....=\mathcal{N}_C=200$, simulation time $=0.042s $, e) group-based: $C= 10 =0.1\%N,  \mathcal{N}_1= \mathcal{N}_2=.....=\mathcal{N}_C=1000$, simulation time $=0.018s $, f) merging of all sub-plots a-e. }      

\label{SIRER}
\end{figure}

The SIR epidemic spreading has three compartments and two types of transitions, one is an edge transition (susceptible to infected) and the other is a nodal transition (infected to recovered). The infected compartment in the influencer compartment of the edge transition. The mean-field approximation of susceptible-infected-recovered (SIR) epidemic model for the individual-based framework are developed by Youssef et. al. \cite{youssef2011individual}. Here, we present the equation for the group-based framework as,

\begin{multline}
\label{SIRExpected1}
  \begin{bmatrix}
 \dot{\rho_{i,S}}\\ \dot{\rho_{i,I}}\\\dot{\rho_{i,R}}
 \end{bmatrix}=  \bigg( \sum_{j=1}^C  \frac{L_{ij}}{\mathcal{N}_i} \rho_
 {j,I}\bigg)\underbrace{\begin{bmatrix} 
 -\beta & \beta &0\\ 0 & 0&0\\0&0&0
 \end{bmatrix}^T}_{Q_\beta^T \text{ matrix}} \begin{bmatrix} \rho_{i,S}\\ \rho_{i,I}\\ \rho_{i,R}\end{bmatrix}\\+ \underbrace{\begin{bmatrix} 
 0 & 0 &0 \\ 0& -\delta & \delta\\ 0 & 0 &0 
 \end{bmatrix}^T}_{Q_\delta^T \text{ matrix}} \begin{bmatrix} \rho_{i,S}\\ \rho_{i,I}\\\rho_{i,R}\end{bmatrix} 
 \end{multline}
 The first part of this equation is for the transition $S\rightarrow I$ and the second part is for the transition $I\rightarrow R$. The number of non linear differential equations for the mean-field approximation of SIR epidemic model for the group-based framework is $(3-1)C$ as at any time $t$, $\rho_{i,S} +\rho_{i,I}+\rho_{i,R}=1$.\\
 The results for the SIR epidemic model in the Erd{\"o}s-R{\'e}nyi are given in Fig. \ref{SIRER}. The network properties and transition rates in the epidemic model are same as the SIS case. The result of stochastic simulation for SIR is presenting in Fig \ref{SIRER}a and mean-filed approximation for individual-based approach is given in Fig \ref{SIRER}b. Group-based approaches are presenting in Fig \ref{SIRER}c-e. The simulation time for individual-based mean-field approach is $12.183s$. The simulation time can be reduced by using group-based approaches. The simulation time reduces with the reduction of the group number ($\leq 0.088s$).

\subsubsection{Susceptible-exposed-infected-recovered}
\begin{figure}[h!]
    \centering
  \centering
    \subfloat[]{\includegraphics[width=1.73in]{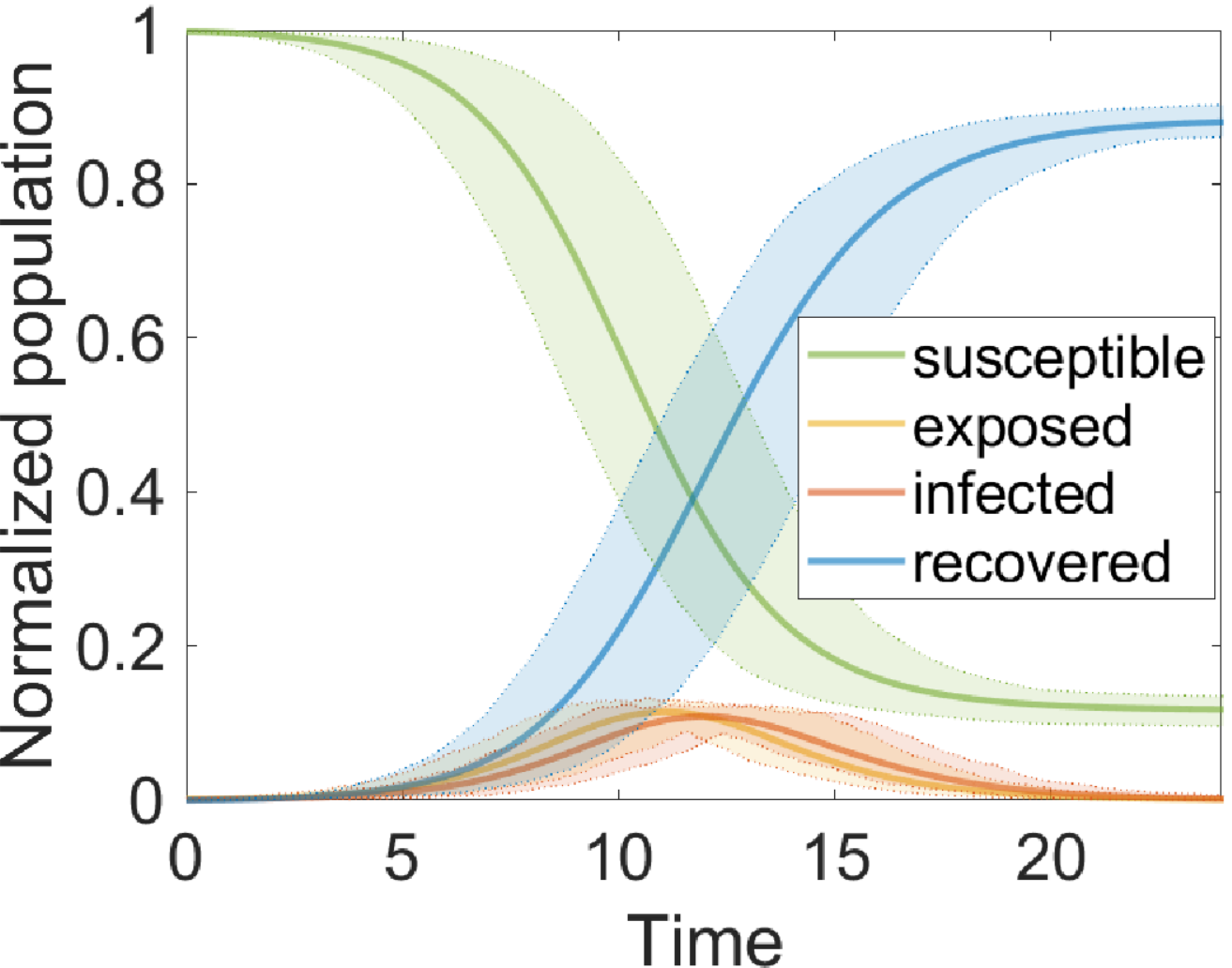}}
    \hfill
    \subfloat[]{\includegraphics[width=1.73in]{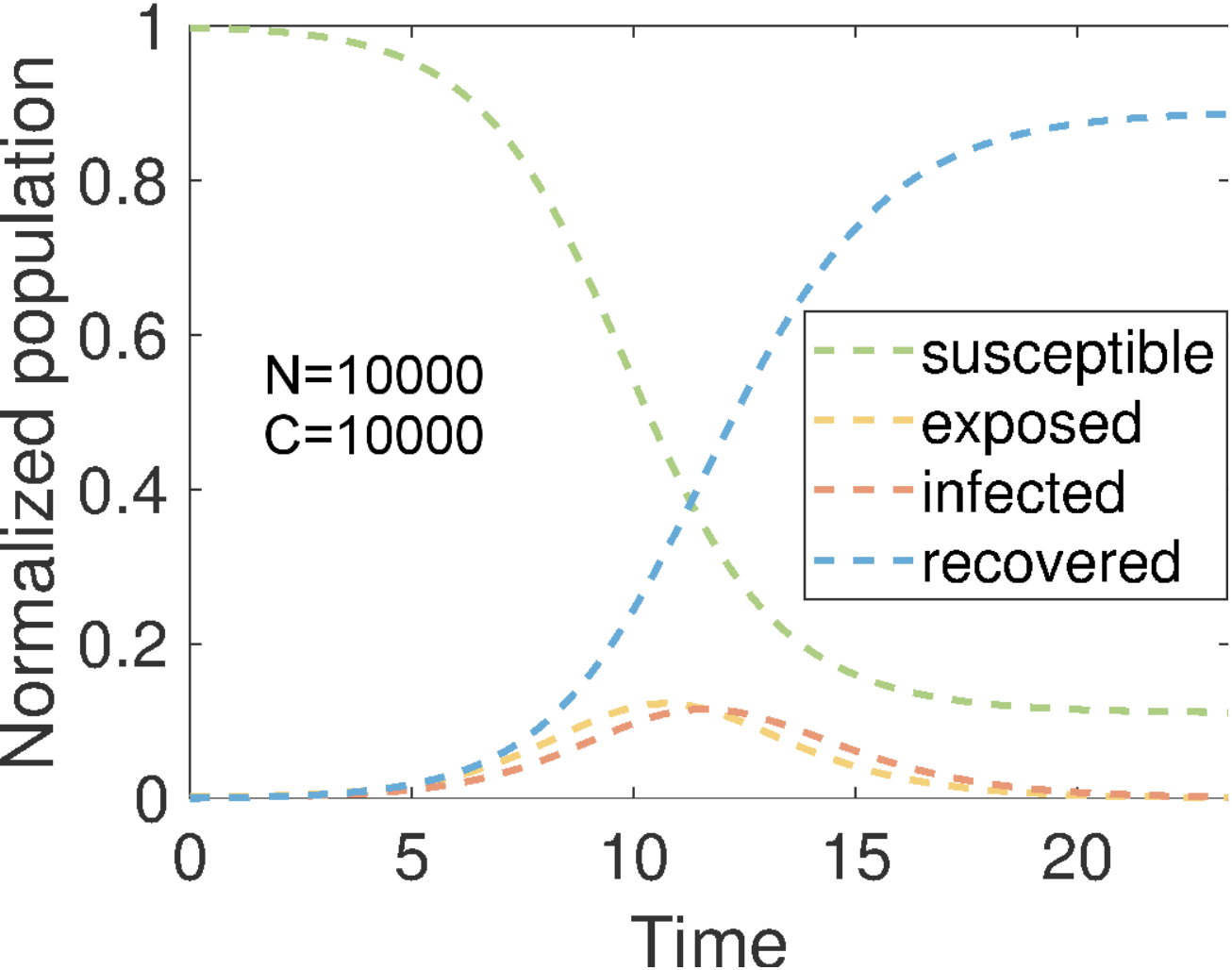}}
   
    \subfloat[]{\includegraphics[width=1.73in]{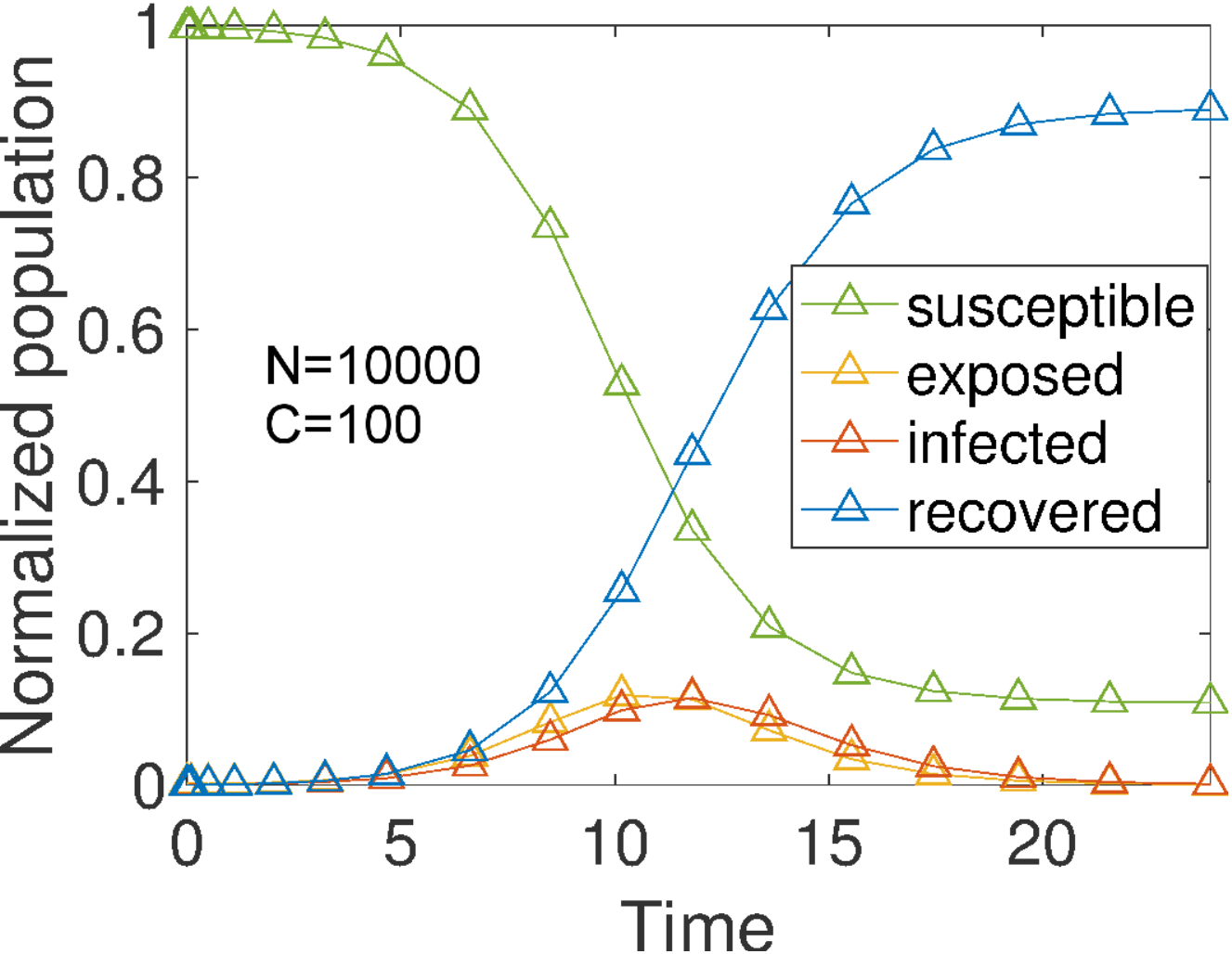}}
      \hfill
    \subfloat[]{\includegraphics[width=1.73in]{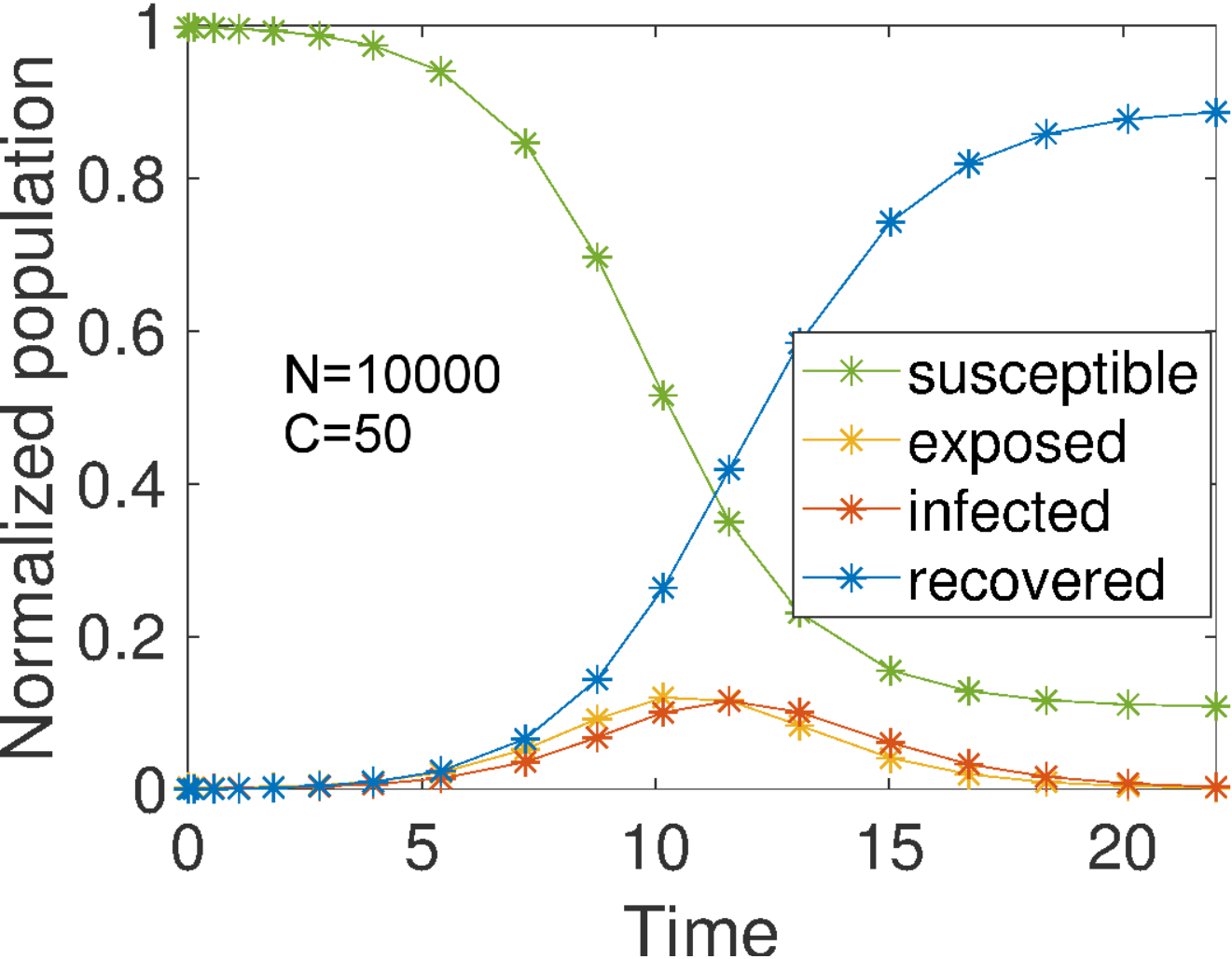}}
    
    \subfloat[]{\includegraphics[width=1.73in]{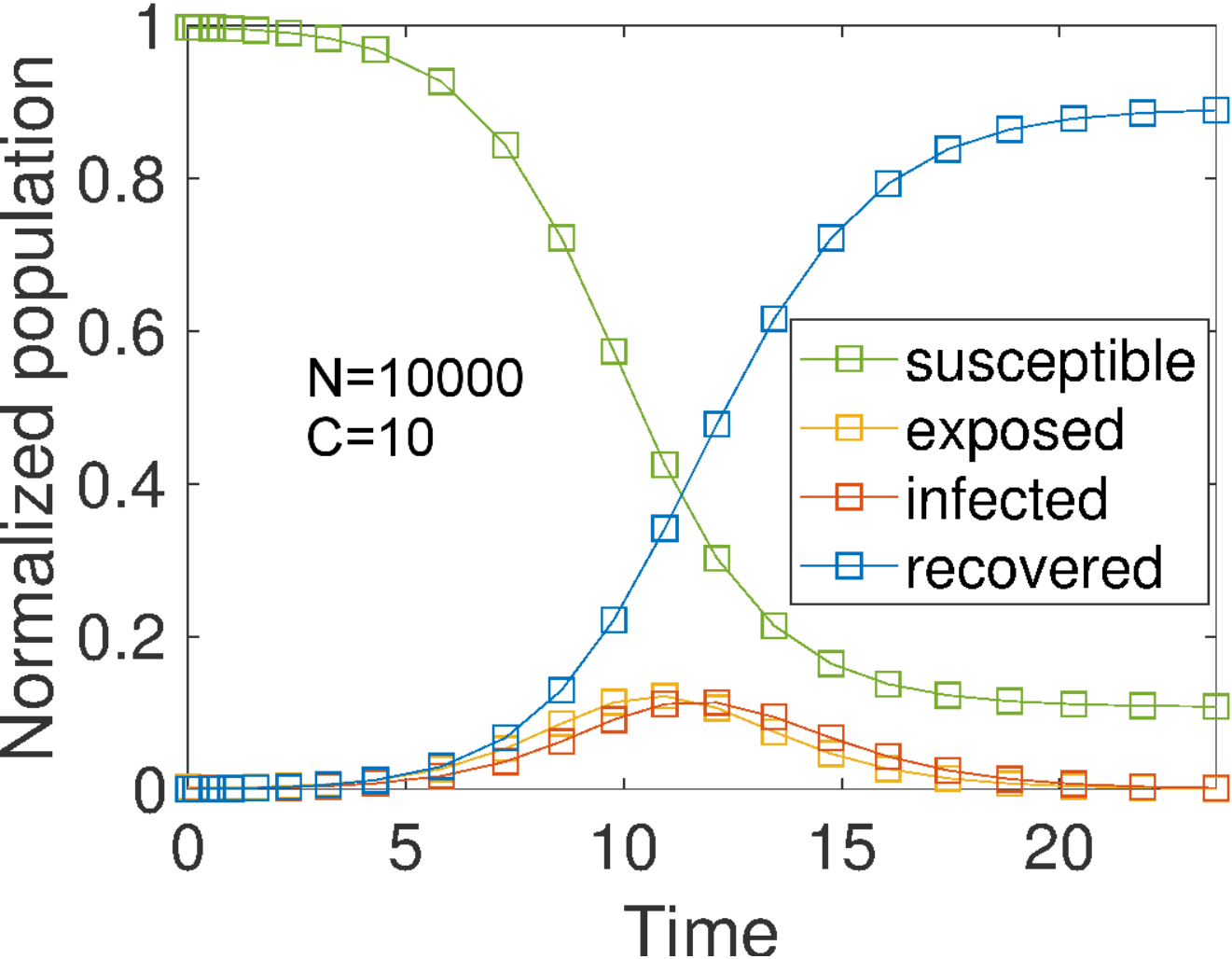}}
      \hfill
    \subfloat[]{\includegraphics[width=1.73in]{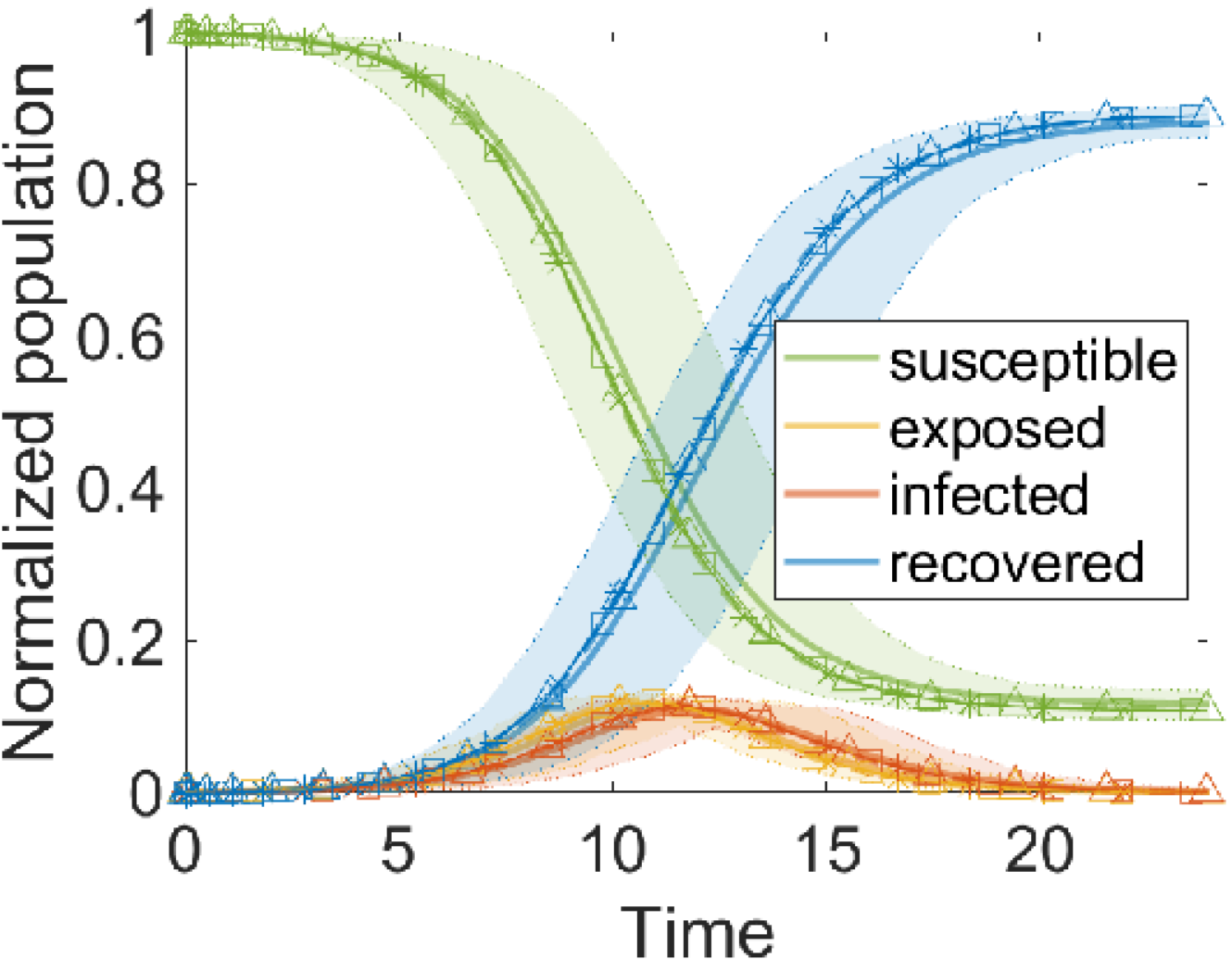}}
       
  \caption{Results for an SEIR epidemic in the Erd{\"o}s-R{\'e}nyi network $(N= 10000, p=0.01)$; a) Stochastic numerical simulation of the Exact Markov process of the individual based approach, solid lines represent the average of the 200 simulations and shaded areas represent region of the stochastic simulation, b) individual-based: $N=C= 10000 , \mathcal{N}_1= \mathcal{N}_2=.....=\mathcal{N}_C=1$, simulation time $=15.264s $, c) group-based: $C= 100 =1\%N ,  \mathcal{N}_1= \mathcal{N}_2=.....=\mathcal{N}_C=100$, simulation time $=0.147s $, d) group-based: $C= 50 =0.5\%N,  \mathcal{N}_1= \mathcal{N}_2=.....=\mathcal{N}_C=200$, simulation time $=0.066s $, e) group-based: $C= 10 =0.1\%N,  \mathcal{N}_1= \mathcal{N}_2=.....=\mathcal{N}_C=1000$, simulation time $=0.019s $, f) merging of all sub-plots a-e. }      

\label{SEIRER}
\end{figure}

The SEIR epidemic model has four compartments and three transitions: susceptible to exposed, exposed to infected and infected to recovered. The first transition is the edge transition, where transition rate ($\beta$) is influenced by the number of infected nodes in the neighboring groups. The other two transitions are nodal transitions with the rate $\delta_1$ ($E \rightarrow I$) and $\delta_2 $ ($I\rightarrow R$) respectively. The group-based mean-field equation for the SEIR epidemic is given below,
\begin{multline}
  \begin{bmatrix}
 \dot{\rho_{i,S}}\\\dot{\rho_{i,E}}\\ \dot{\rho_{i,I}}\\\dot{\rho_{i,R}}
 \end{bmatrix}= \bigg( \sum_{j=1}^C  \frac{L_{ij}}{\mathcal{N}_i} \rho_{j,I}\bigg)\underbrace{\begin{bmatrix} 
 -\beta & \beta &0&0\\ 0 & 0&0&0\\0&0&0&0\\0&0&0&0
 \end{bmatrix}^T}_{Q_\beta^T \text{ matrix}} \begin{bmatrix} \rho_{i,S}\\\rho_{i,E}\\ \rho_{i,I}\\ \rho_{i,R}\end{bmatrix}\\+\underbrace{\begin{bmatrix} 
 0 & 0 &0 &0\\ 0& -\delta_1 & \delta_1&0\\ 0 & 0 &0 &0\\0 & 0 &0 &0
 \end{bmatrix}^T}_{Q_{\delta_1}^T \text{ matrix}} \begin{bmatrix} \rho_{i,S}\\\rho_{i,E}\\ \rho_{i,I}\\\rho_{i,R}\end{bmatrix} +\underbrace{\begin{bmatrix} 
 0 & 0 &0 &0\\ 0& 0 &0&0\\ 0 & 0 &-\delta_2 &\delta_2\\0 & 0 &0 &0
 \end{bmatrix}^T}_{Q_{\delta_2}^T \text{ matrix}} \begin{bmatrix} \rho_{i,S}\\\rho_{i,E}\\ \rho_{i,I}\\\rho_{i,R}\end{bmatrix} 
 \label{mfel}
 \end{multline}
 The first part of the Eq. (\ref{mfel}) represents the edge transition $S \rightarrow E$, the second part represents the nodal transition $E \rightarrow I$ and the last part represents the nodal transition $I \rightarrow R$. The number of nonlinear ordinary differential equations for this epidemic model is $(4-1)C$, as at any time $t$, $\rho_{i,S}+\rho_{i,E}+\rho_{i,I}+\rho_{i,R}=1$. The results for the SEIR are given in Fig. \ref{SEIRER}. Here, the value of $\beta$, $\delta_1$, and $\delta_2$ is 0.025, 1, and 1 accordingly. Network properties and initial condition are same as the SIS epidemic model.

A summary of Fig \ref{SISER}, \ref{SIRER}, and \ref{SEIRER} is given in the following table. Computational environment was same for each case. 
\begin{table}[htb]
\renewcommand{\arraystretch}{1.5}
\caption{}
\label{table_mfa}
\centering
\begin{tabular}{  p{0.35\linewidth}>{\centering\arraybackslash} p{0.09\linewidth}>{\centering\arraybackslash} p{0.09\linewidth}>{\centering\arraybackslash} p{0.09\linewidth}>{\centering\arraybackslash}  p{0.09\linewidth}}
\hline
\hline
case & No. of Groups & & simulation time&\\
\hline
\hline
& & SIS&SIR &SEIR\\
\hline
Individual-based stochastic & -&2140s &  348s & 595s \\
Individual-based mean-field& -&10.409s & 12.183s&15.264s\\
group-based mean-field & 100&0.084s & 0.088s&0.147s\\
group-based mean-field & 50&  0.049s& 0.042s&0.066s\\
group-based mean-field & 10& 0.016s& 0.018s&0.019s\\
\hline
\end{tabular}
\end{table}

The group-based approach is a solution with reduced computational time. However, because of topological and moment closure approximation, the results can deviate from the exact process for some scenarios. It is not the scope of this paper. From previous research works, mean-field SIS model is less accurate in the sparse graphs \cite{sahneh2013generalized}. The accuracy of the mean-field models is also sensitive to the network structure. The mean-field model can follow the exact process very closely when the size of the network is very large \cite{li2012susceptible}. An extensive numerical simulation of GgroupEM in different scenario with respect to different network structures, initial conditions, and group sizes can be a valuable research topic for future analysis.

\section{Multilayer extension of the GgroupEM}
\label{Multilayerextension}
\begin{figure}[ht]
    \centering
  \includegraphics[width=\linewidth]{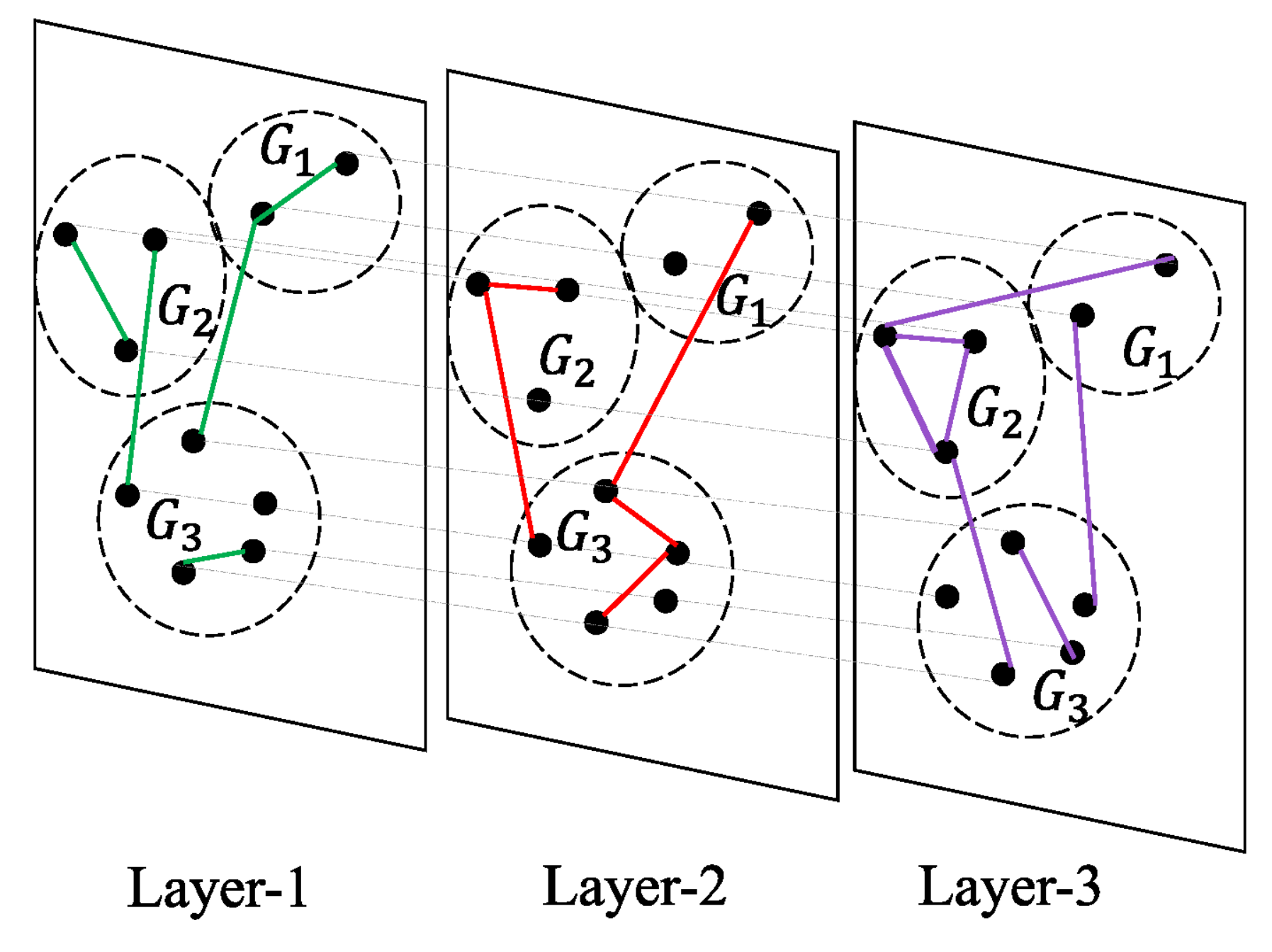}
       
  \caption{Example of a multilayer network, which has three layers. The nodes are divided into three groups, $G-1, G-2, \text{ and } G-3$ . }      

\label{multilayer}
\end{figure}

In the real world, contact network among interacting agents can have a complex structure, where the nature of the connection between two agents can be of multiple types. For example, in the rumor spreading network, two people can be connected via Facebook or they can be connected via Twitter. To represents these complex structures, researchers are using multilayer networks \cite{kivela2014multilayer, kurant2006layered}, where each layer represents each type of connection. If a social network has three types of connections: direct connection, Facebook connection, and twitter connection, then a three-layer network can be used to represent this network where each layer corresponds to each type of connections. In a disease spreading network, if a disease spreads through direct contact and by air, then a two-layer network will represent the network more precisely; one layer is for direct contact and another layer is for the air transmission. \\In the group-based structure, nodes and groups will be maintained in each layer, however, the connection among them will be different for different layers. An example of a group-based multilayer network, presented in Fig. \ref{multilayer} has three layers. Groups are the same for each layer, however, the connections are different for each layer. In particular, green lines form the link of layer-1, red lines form the link of layer-2, and purple lines form the link of layer-3.

If the network has $\mathcal{L}$ layers, then the Eq. (\ref{differetiale}) can be modified as,
\begin{multline}
  \frac{d}{dt}E[g_i(t)]=   \sum_{ q=1}^{q_n}{ (\Delta_{i,\delta_q}^T\circ \mathcal{V}_{i,m})E[g_i(t)]} \\+ \sum_{ q=1}^{q_e}{   \bigg(\sum_{l=1}^{\mathcal{L}}(\Delta_{i,\beta_{ql}}^T\circ \mathcal{V}_{i,m})^TE[h_{il}(t)g_i(t)]}\bigg)
\end{multline}
Here, the matrix for edge transition $\Delta_{i,\beta_{ql}}$ is layer-specific, and $ h_{il}(t)=\sum_{j=1}^{C}A^{gl}_{ij}X_{j,n}$. The transition rate for edge transition can different for different layers. Also, the mean field equation Eq. (\ref{mf equation}) can be modified for the multilayer network as,
\begin{multline}
\label{multiLayerequation}
    \frac{d}{dt}E[X_i]=\sum_{ q=1}^{q_n}Q_{\delta_q}^TE[X_i] \\+\sum_{q=1}^{q_e}\bigg( \sum_{l=1}^{\mathcal{L}}\big(\sum_{j=1}^C A^{gl}_{ij} E[X_{j,n}]\big)Q_{\beta_{ql}}^T\bigg)E[X_i]
\end{multline}
Only the parts for edge transition in Eq. (\ref{differetiale}) and (\ref{mf equation}) are needed to be modified for the multilayer extension, as nodal transitions are independent of the network structure. The Eq. (\ref{mfe1})-(\ref{mfel}) can be rewritten in this similar manners as Eq. (\ref{multiLayerequation}).
\section{Conclusion}
\label{Conclusion}
In this paper, we propose a general group-based epidemic model (GgroupEM) framework capable of representing any compartmental model in any multilayer networks. \\
We develop a continuous-time Markov model for the group-based approach that has  $\binom{\mathcal{N}_1 +M-1}{M-1}\binom{\mathcal{N}_2 +M-1}{M-1}....\binom{\mathcal{N}_C +M-1}{M-1}$ possible states. This is a multidimensional birth-death process. The possible states in the Markov chain of the GgroupEM are fewer than or equal to the possible states in the Markov chain of the individual-based approach, which are $M^N$. Therefore, GgroupEM has a reduced computational complexity and it requires less simulation time. Simulation time is important when considering very large networks. 
 The group-based process lies on an approximation based on the isoperimetric inequality. We further reduced the number of states by using a moment-closure approximation. The $N$-intertwined mean-field approximation (NIMFA) method \cite{chakrabarti2008epidemic, van2009virus} and the heterogeneous mean-field method (HMF) \cite{pastor2001epidemic, boguna2002epidemic} are two well-known methods of the moment-closure approximation, which are two particular cases of the group-based mean-field method. The number of the nonlinear differential equation for the mean-field approximation of the group-based approach is $(M-1)C$. Finally, we present some simulation results of the mean-field approximation for SIS, SIR, and SEIR epidemic model in the Erd{\"o}s-R{\'e}nyi random network. For each case, we find that simulation time reduces with the reduction of the number of groups.\\
 The GgroupEM framework lies on two approximations; topological approximation and moment-closure approximation. The topological approximation is for the underlying network and the error for this approximation can be bounded by the isoperimetric inequality. On the other hand, for the moment closure approximation, we only know that for the $C=N$ grouping, the moment-closure approximation is the upper bound of the exact process. However, we do not have exact knowledge about the error bound for the moment-closure approximation. The accuracy of the mean-field model has been explained in \cite{sahneh2013generalized,van2009virus}.
\\The group-based approach allows us to scale the network and reduce computational time. It is possible to obtain the disease dynamics of an epidemic model in a large complex network by using GgroupEM when aggregated dynamics of groups of nodes are the focus of interest.\\
In this paper, we have developed a general group-based framework (GgroupEM). An extensive performance analysis of the group-based approach in different types of networks with different initial conditions can be an interesting future step of this research. A development of a continuous-time numerical stochastic simulator for GgroupEM will allow the researchers to compare GgroupEM with GEMFsim \cite{sahneh2017gemfsim}.


%

\appendices
\numberwithin{equation}{section}

\section{Derivation of the $\Theta$}
\label{appendixEMP}

In this section, we present the derivation of the $\Theta$ in Eq. (\ref{exacte}). To do this, we derive the expression for network state $G(t +\Delta t)$ when $G(t)$ is given. Here, $\Delta t$ is a very small time period when only one event can occur. 
Let, the network state at any time $t$ is,
\begin{equation}
    G(t)= g_Z= g_{z1}(t)\otimes...........\otimes g_{zC}(t)
\end{equation}
The network state will change by the one transition in the group state of group $i$,
\begin{multline}
    E[G(t+\Delta t)|G(t)= g_Z]= \sum_{i=1}^Cg_{z1}(t)\otimes .........\otimes\\ E[g_i(t+\Delta t)|G(t)=g_Z]\otimes...\otimes g_{zC}(t)
\end{multline}
The expression for the conditional expectation of a group $E[g_i(t+\Delta t)|G(t)=g_Z]$ can get from the Eq. (\ref{onesideEquation}) as, 
\begin{multline}
    E[g_i(t+\Delta t)|G(t)=g_Z]=  \sum_{ q=1}^{q_n}{ (\Delta_{i,\delta_q}^T\circ \mathcal{V}_{i,m})g_{zi}(t)}\Delta t \\+ \sum_{q=1}^{q_e}{\bigg(\sum_{j=1}^{C}\mathcal{A}_g(i,j)X_{j,n}\bigg) (\Delta_{i,\beta_q}^T\circ \mathcal{V}_{i,m})g_{zi}(t)}\Delta t \\+ g_{zi}(t) + o(\Delta t)
    \label{foronege}
\end{multline}

Now, from the definition of expectation and the law of total probability we get the network state at $t + \Delta t$ time,
\begin{multline}
    E(G(t+\Delta t)) = \sum_{Z}{E[G(t +\Delta t)|G(t)=g_Z]Pr[G(t)= g_Z]}
    \label{forallze}
\end{multline}
Here, the range of Z for the summation in Eq. (\ref{forallze}) is $1 : \binom{\mathcal{N}_1 +M-1}{M-1}\binom{\mathcal{N}_2 +M-1}{M-1}....\binom{\mathcal{N}_C +M-1}{M-1}$.\\
From Eq. (\ref{foronege}) and (\ref{forallze}), 

\begin{multline}
    E(G(t+\Delta t)) =  \sum_{ q=1}^{q_n}{ \Theta_{\delta_q}E[G(t)]\Delta t} \\+ \sum_{ q=1}^{q_e}{\Theta_{\beta_q}E[G(t)]\Delta t} + E[G(t)]+o(\Delta t)
    \label{contiMardifferebce}
\end{multline}\
Here, 
\begin{multline}
     \Theta_{\delta_q}= \sum_{i=1}^CI_{\binom{\mathcal{N}_1 +M-1}{M-1} \times \binom{\mathcal{N}_1 +M-1}{M-1}}\otimes....\otimes (\Delta_{\delta_q, i}^T\circ \mathcal{V}_{i,m}).\\...\otimes I_{\binom{\mathcal{N}_C +M-1}{M-1} \times \binom{\mathcal{N}_C +M-1}{M-1}}
     \label{thetadelta}
\end{multline}
The Z\textsuperscript{th} column of $\Theta_{\beta_q}$ is,
\begin{multline}
    \Theta_{\beta_q}(:,Z)=\sum_{i=1}^Cg_{z1}(t)\otimes..\\...\otimes \bigg(\sum_{j=1}^{C}\mathcal{A}_g(i,j)X_{j,n}\bigg) (\Delta_{i,\beta_q}^T\circ \mathcal{V}_{i,m})g_{zi}(t) \otimes....\otimes g_{zC}(t)
    \label{thetabeta}
\end{multline}
Let,
\begin{equation}
    \Theta=\sum_{ q=1}^{q_n}\Theta_{\delta_q} +\sum_{q=1}^{q_e} \Theta_{\beta_q}
\end{equation}
The differential equation for the underlying continuous-time Markov process for the group-based approach we will get from Eq. (\ref{contiMardifferebce}) by letting $\Delta t \rightarrow 0$,
\begin{equation}
    \frac{d}{dt}E[G]= \Theta E[G]
\end{equation}
\section{An example of the group-based epidemic model }
\label{Appendixexample}

A network with $N=5$ nodes. The nodes are divided into $C=2$ groups.First group has $\mathcal{N}_1 =2$ nodes and second group has $\mathcal{N}_2 =3$. For susceptible-infected-susceptible (SIS) epidemic process the first group has $\binom{\mathcal{N}_1 +M-1}{M-1}= 3$ states and the second group has  $\binom{\mathcal{N}_2 +M-1}{M-1}= 4$ states. The description of the group states are given below, the left matrix is for group-1 and the right matrix is for group-2.
\begin{multline}
\begin{split}
    group_1=\begin{bmatrix}
 \_ & \_&\_&V_1\\
 | &o&o&[0,2] \\
 o&|&o&[1,1] \\
 o&o&1&[2,0]
\end{bmatrix} \\
group_2=\begin{bmatrix}
 \_ & \_& \_&\_&V_2\\
 | &o&o&o&[0,3] \\
 o &|&o&o&[1,2] \\
 o &o&|&o&[2,1] \\
 o &o&o&|&[3,0] 
\end{bmatrix}
\end{split}
\end{multline}
SIS epidemic process has two compartments; susceptible and infected. One divider $|$ can divides the nodes $o$ into two compartments. At first, we will present the steps to get $\Theta_{\delta_1}$ for the nodal transition infected to susceptible compartment, then we will present the steps to get $\Theta_{\beta_1}$ for the edge transition susceptible to infected compartment.\\
Here, the \textit{transition indication matrix} for the group-1 and group-2 for the nodal transition from susceptible to infected compartment will be,
\begin{multline}
\begin{split}
   \Delta_{1,\delta_1}= \begin{bmatrix} 
   -\delta & \delta & 0\\
   0 & -\delta & \delta\\
   0 & 0 &0
    \end{bmatrix}\\
  \Delta_{2,\delta_1}  =\begin{bmatrix}
  -\delta & \delta &0 &0\\
  0 & -\delta & \delta &0\\
  0&0 &-\delta & \delta\\
  0& 0& 0 &0
  \end{bmatrix}
  \end{split}
  \label{exampletransitionindication}
\end{multline}
The definition of the \textit{transition indication matrix} is given in Eq. (\ref{transitionImatrix}).\\
Now, $\mathcal{V}_{i,m}$ in Eq. for the infected (compartment $m=2$) to susceptible (compartment $m=1$) nodal transition will be,
\begin{multline}
\begin{split}
   \mathcal{V}_{1,2}= \begin{bmatrix} 
   2 & 1 & 0\\
   2 & 1 & 0\\
   2 & 1 & 0
    \end{bmatrix}\\
   \mathcal{V}_{2,2}  =\begin{bmatrix}
  3 & 2&1&0\\
  3 & 2&1&0\\
  3 & 2&1&0\\
  3 & 2&1&0
  \end{bmatrix}
  \end{split}
  \label{examplevim1}
\end{multline}
Then,
\begin{multline}
\begin{split}
\Delta_{1,\delta_1}^T\circ \mathcal{V}_{1,2}= \begin{bmatrix} 
-2\delta &0&0\\
2\delta&-\delta&0\\
0& \delta&0
\end{bmatrix}\\
\Delta_{2,\delta_1}^T\circ \mathcal{V}_{2,2}= \begin{bmatrix} 
-3\delta & 0 &0 &0\\
  3\delta & -2\delta & 0 &0\\
  0& 2\delta &-\delta & 0\\
  0& 0& \delta &0
\end{bmatrix}
\end{split}
\end{multline}
\begin{figure*}[!h]
\normalsize
\begin{multline}
\label{thetadelta1}
\Theta_{\delta_1}= \begin{bmatrix}
-5\delta & 0       & 0      & 0      & 0 & 0 & 0 & 0 & 0 & 0 & 0 & 0\\
3\delta  &-4\delta & 0      & 0      & 0 & 0 & 0 & 0 & 0 & 0 & 0 & 0\\
0        &2\delta  &-3\delta& 0      & 0 & 0 & 0 & 0 & 0 & 0 & 0 & 0\\
0        &  0      &\delta  &-2\delta& 0 & 0 & 0 & 0 & 0 & 0 & 0 & 0\\
2\delta&0&0&0&-4\delta&0&0&0&0&0&0&0\\
0&2\delta&0&0&3\delta&-3\delta&0&0&0&0&0&0\\
0&0&2\delta&0&0&2\delta&-2\delta&0&0&0&0&0\\
0&0&0&2\delta&0&0&\delta&-\delta&0&0&0&0\\
0&0&0&0&\delta&0&0&0&-3\delta&0&0&0\\
0&0&0&0&0&\delta&0&0&3\delta&-2\delta&0&0\\
0&0&0&0&0&0&\delta&0&0&2\delta&-\delta&0\\
0&0&0&0&0&0&0&\delta&0&0&\delta&0
\end{bmatrix}
\end{multline}
\end{figure*}

The $\Theta_{\delta_1}$ matrix can get for this case from Eq. (\ref{thetadelta}), which is presented in Eq (\ref{thetadelta1}).\\ 
let, the states of the groups at time $t$ are, $g_{z1}(t)=\begin{bmatrix}0\\1\\0\end{bmatrix}$, and $g_{z2}(t)=\begin{bmatrix}0\\0\\1\\0\end{bmatrix}$. Therefore, the group state at time $t$ is $G(t)=\begin{bmatrix}0,0,0,0,0,0,1,0,0,0,0,0\end{bmatrix}^T$.\\
Now, the \textit{transition indication matrix} for the edge transition from susceptible to infected compartment with the rate $\beta$ is,
\begin{multline}
\begin{split}
\Delta_{1,\beta_1}= \begin{bmatrix} 
   0 & 0 & 0\\
   \beta & -\beta & 0\\
   0 & \beta & -\beta
    \end{bmatrix}\\
  \Delta_{2,\beta_1}  =\begin{bmatrix}
  0 & 0 &0 &0\\
  \beta & -\beta & 0 &0\\
  0&\beta & -\beta & 0\\
  0& 0& \beta & -\beta
  \end{bmatrix}
  \end{split}
\end{multline}
Now, $\mathcal{V}_{i,m}$ for the edge transition susceptible (compartment $m=1$) to infected (compartment $m=2$) will be,
\begin{multline}
\begin{split}
   \mathcal{V}_{11}= \begin{bmatrix} 
   0 & 1 & 2\\
   0 & 1 & 2\\
   0 & 1 & 2
    \end{bmatrix}\\
   \mathcal{V}_{21}  =\begin{bmatrix}
  0 & 1&2&3\\
  0 & 1&2&3\\
  0 & 1&2&3\\
  0 & 1&2&3
  \end{bmatrix}
  \end{split}
  \label{examplevim2}
\end{multline}
Therefore,
\begin{multline}
\begin{split}
\Delta_{1,\beta_1}^T\circ \mathcal{V}_{11}= \begin{bmatrix} 
0 &\beta&0\\
0&-\beta&2\beta\\
0& 0&-2\beta
\end{bmatrix}\\
\Delta_{2,\beta_1}^T\circ \mathcal{V}_{21}= \begin{bmatrix} 
0 & \beta &0 &0\\
  0 & -\beta& 2\beta &0\\
  0&0 &-2\beta & 3\beta\\
  0& 0& 0 &-3\beta
\end{bmatrix}
\end{split}
\end{multline}
Now, the $\Theta _{\beta_1}$ matrix for this case can get from Eq. (\ref{thetabeta}),
\begin{equation}
\Theta_{\beta_1}(:,Z)=\begin{bmatrix}
0\\
0\\
\beta \bigg(\sum_{j=1}^{2}{\mathcal{A}_g(1,j)X_{j,2}}\bigg)\\
0\\
0\\
2\beta \bigg(\sum_{j=1}^{2}\mathcal{A}_g(2,j)X_{j,2}\bigg)\\
-2\beta \bigg(\sum_{j=1}^{2}\mathcal{A}_g(2,j)X_{j,2}\bigg)- \beta \bigg(\sum_{j=1}^{2}\mathcal{A}_g(1,j)X_{j,2}\bigg)\\
0\\
0\\
0\\
0\\
0
\end{bmatrix}
\end{equation}
\section*{Acknowledgment}

Authors would like to express their gratitude to Piet Van Mieghem, Karel Devriendt, and Bastian Prasse for useful insights. This work was supported by the NSF/NIH/USDA/BBSRC Ecology and Evolution of Infectious Diseases (EEID) Program through USDA-NIFA Award 2015-67013-23818.

\ifCLASSOPTIONcaptionsoff
  \newpage
\fi



%



\bibliographystyle{IEEEtran}
\bibliography{eee}

%

\begin{IEEEbiography}[{\includegraphics[width=1in,height=1.25in,clip,keepaspectratio]{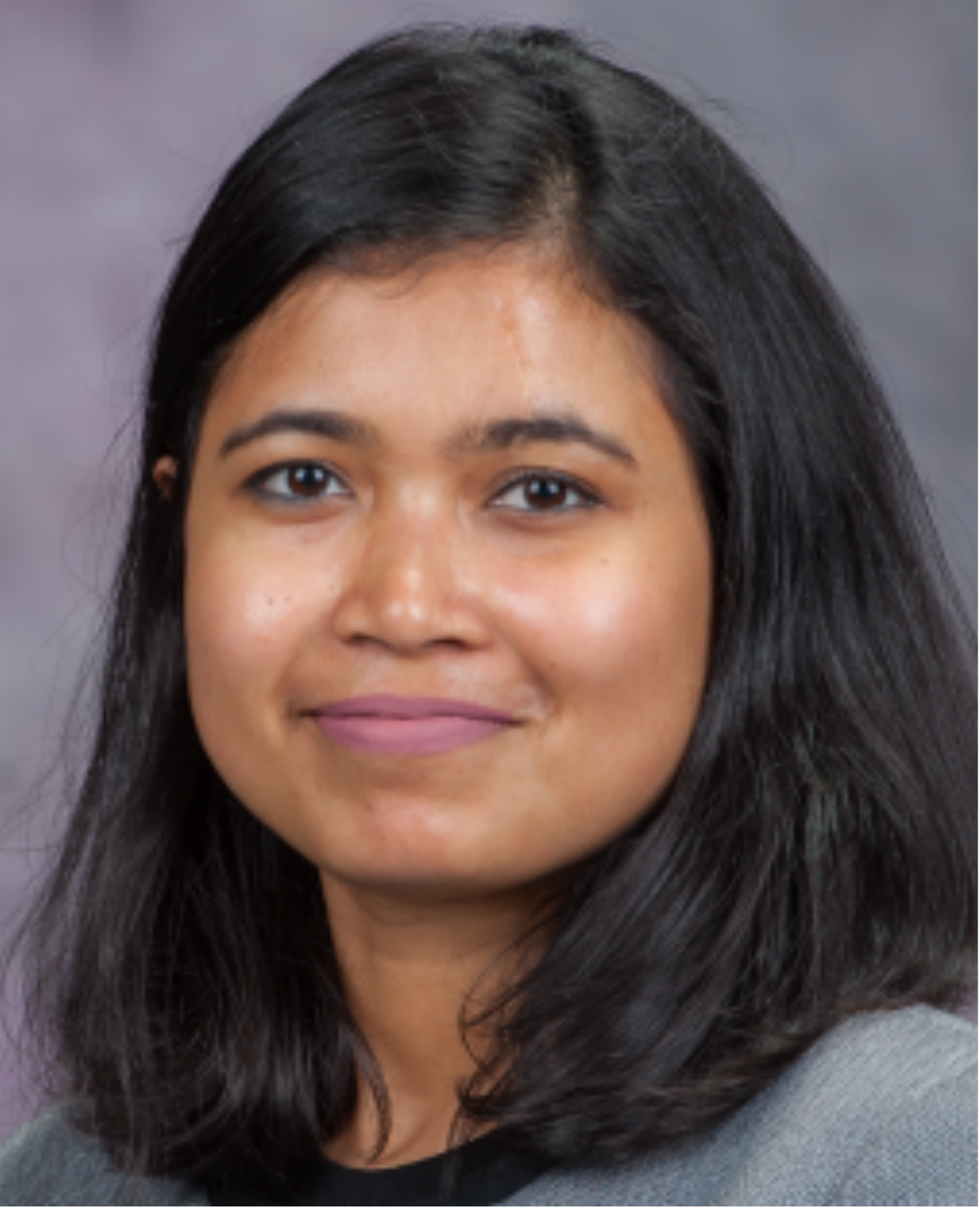}}]{Sifat Afroj Moon}
graduated with a degree in Electrical and Electronic 
Engineering (EEE) from the Bangladesh University of Engineering and Technology (BUET) in 2013. From 2013 to 2016, she works as a software engineer in the Samsung research and development center. She has started her Ph.D degree at the Kansas State University in the computer engineering under the supervision of Dr. Caterina Scoglio in 2016.\\
Her research interest in on the network science, modeling of Spreading process on complex networks, and developing the network based technology or tools.
\end{IEEEbiography}

\begin{IEEEbiography}[{\includegraphics[width=1in,height=1.25in,clip,keepaspectratio]{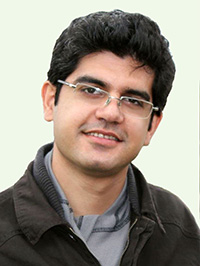}}]{Faryad Darabi Sahneh} received the B.S. degree in mechanical engineering from Amirkabir University of Technology (Tehran Polytechnic), Tehran, Iran, in 2008, the M.S. degree in mechanical engineering from Kansas State University, Manhattan, KS, USA, in 2010, and the Ph.D. degree in electrical and computer engineering from Kansas State University. Then he worked as a postdoc in Georgia Institute of Technology for one years. Currently, he is working with the department of mathematics in the University of Arizona, Tucson, AZ, USA. 
\end{IEEEbiography}


\begin{IEEEbiography}[{\includegraphics[width=1in,height=1.25in,clip,keepaspectratio]{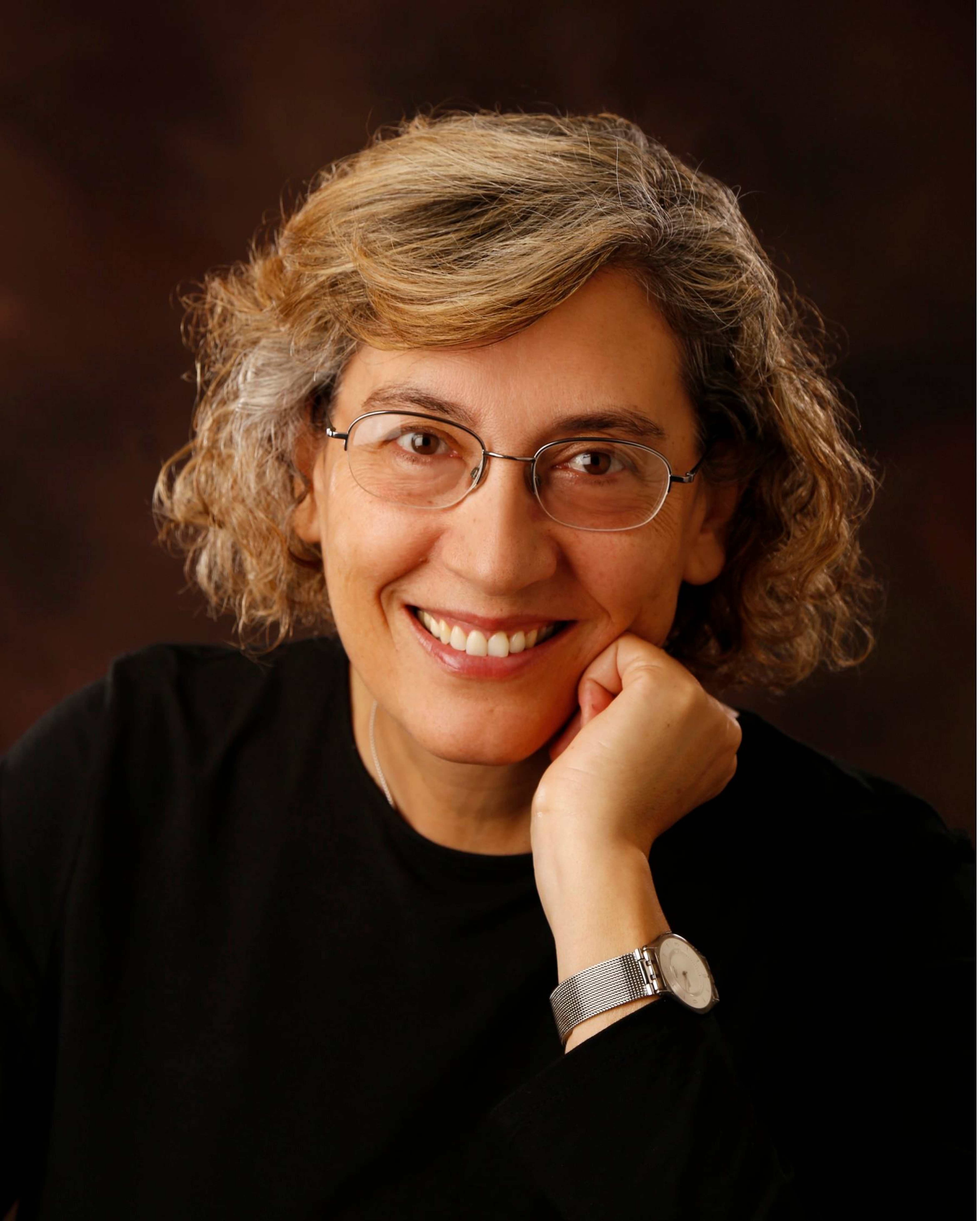}}]{Caterina Scoglio} is the Paslay chair professor of Electrical and Computer Engineering at Kansas State University. Her main research interests are in the field of network science and engineering. Caterina received the Dr. Eng. degree from the "Sapienza" Rome University, Italy, in 1987. Before joining Kansas State University, she worked at the Fondazione Ugo Bordoni from 1987 to 2000, and at the Georgia Institute of Technology from 2000 to 2005. Caterina is also affiliated faculty member at the Institute of Computational Comparative Medicine (ICCM) at Kansas State University.
\end{IEEEbiography}




\end{document}